\PassOptionsToPackage{unicode}{hyperref}
\PassOptionsToPackage{hyphens}{url}

\documentclass[sigplan,screen,10pt]{acmart}
\acmConference[Just a Draft]{Just a Draft}{Just a Draft}{Just a Draft}
\acmBooktitle{Just a Draft}
\acmPrice{Just a Draft}
\acmISBN{Just a Draft}

\usepackage{hyperref}
\usepackage{amsmath}
\usepackage{cleveref}
\usepackage{adjustbox}

\usepackage{macros}

\usepackage{mathtools}
\DeclarePairedDelimiter\ceil{\lceil}{\rceil}
\DeclarePairedDelimiter\floor{\lfloor}{\rfloor}

 \usepackage{xspace}
 \usepackage{graphicx}
 \usepackage{adjustbox}
 \usepackage{dsfont}
 \usepackage[colorinlistoftodos]{todonotes}
 \usepackage[inline]{enumitem}
 \setlist[itemize]{topsep=0pt,partopsep=0pt,itemsep=0pt,parsep=0pt}
 \setlist[itemize,1]{label=-}
 \setlist[itemize,2]{label=---}
 \setlist[itemize,3]{label=*}
 \setlist[enumerate]{topsep=0pt,partopsep=0pt,itemsep=0pt,parsep=0pt}
 \setlist[enumerate,1]{label=\roman*)}
 \setlist[enumerate,2]{label=\alph*)}
 \setlist[enumerate,3]{label=\arabic*)}
 \usepackage{multicol}
 \usepackage{xcolor}
 
\usepackage{tikz}
\usetikzlibrary{arrows.meta}
\usetikzlibrary{shapes}
\usetikzlibrary{calc}
\usetikzlibrary{math}
\usetikzlibrary{decorations.pathreplacing,calligraphy}
\usepackage{pgffor}

\usepackage{algorithm}
\usepackage{algpseudocode}

\newcommand{\peer}[1]{\ensuremath{\mathcal{X}_{#1}}}

\newcommand{\examplea}[0]{\ensuremath{\mathrm{ape}}}
\newcommand{\exampleb}[0]{\ensuremath{\mathrm{bee}}}
\newcommand{\examplec}[0]{\ensuremath{\mathrm{cat}}}
\newcommand{\exampled}[0]{\ensuremath{\mathrm{doe}}}
\newcommand{\examplee}[0]{\ensuremath{\mathrm{eel}}}
\newcommand{\examplef}[0]{\ensuremath{\mathrm{fox}}}
\newcommand{\exampleg}[0]{\ensuremath{\mathrm{gnu}}}
\newcommand{\exampleh}[0]{\ensuremath{\mathrm{hog}}}
\newcommand{\examplei}[0]{\ensuremath{\examplea}}

\newcommand{\return}[0]{\textbf{return }}

\definecolor{leftSearch}{HTML}{428BFF}
\definecolor{rightSearch}{HTML}{FC4842}
\definecolor{leftInner}{HTML}{AA42FF}
\definecolor{rightInner}{HTML}{F79C40}
\definecolor{top}{HTML}{F5FF3D}

\tikzset{
aux/.style={
  align=left,
  draw=black,
  fill=white
  }
}

\newcommand{\examplefpi}[3]{
  $\ifpmanual{#1}{#2}{\fp{#3}}$
}
\newcommand{\exampleiis}[4]{
\begin{tabular}{ c | c  | c }
{#1} & {#3} & {#2} \\ 
\end{tabular} $_{(#4)}$
}

\tikzset{
fpi/.style={
  align=left,
  draw=black,
  fill=white
  }
}

\tikzset{
iis/.style={
  align=left,
  draw=black,
  fill=white,rounded corners=.25cm
  }
}

\tikzset{
local/.style={
  }
}

\tikzstyle{edge} = [draw,thick,opacity=0.25]
\tikzstyle{dep} = [draw,thick]

\tikzset{
skiplistnode/.style={
  align=center,
  draw=black,
  fill=white,
minimum width={width("\exampleg") + width("x")},
minimum height={height("\exampleh") + height("h"}
  }
}

\tikzset{
fingerprintnode/.style={
  align=center,
  draw=black,
  thick,
  fill=lightgray,
minimum width={width("\exampleg") + width("x")},
minimum height={height("\exampleh") + height("h"}
  }
}

\title{Range-Based Set Reconciliation}

\author{Aljoscha Meyer}
\begin{document}

\begin{abstract}
Range-based set reconciliation is a simple approach to efficiently computing the union of two sets over a network, based on recursively partitioning the sets and comparing fingerprints of the partitions to probabilistically detect whether a partition requires further work. Whereas prior presentations of this approach focus on specific fingerprinting schemes for specific use-cases, we give a more generic description and analysis in the broader context of set reconciliation. Precisely capturing the design space for fingerprinting schemes allows us to survey for cryptographically secure schemes. Furthermore, we reduce the time complexity of local computations by a logarithmic factor compared to previous publications.
\end{abstract}

\maketitle

\textbf{This draft has not been peer-reviewed!}

Ignore the ACM copyright section, this is a public-domain draft that just happens to use an ACM template.

\section{Introduction}\label{introduction}

Set reconciliation is the problem of computing the union of two sets that are located at two different nodes in a network; both nodes should hold the union of the two sets after reconciling. If both nodes simply send their full set, the potentially large intersection of the two sets is transmitted redundantly. Hence we are interested in (probabilistic) solutions whose communication complexity can be bounded by the size of the symmetric difference of the two sets.

An archetypical use case for set reconciliation occurs in peer-to-peer systems for information sharing. In such systems, nodes connect to randomly chosen other nodes, each supplying missing data to their peers. The nodes often have restrictive computational capabilities, yet they need to perform frequent reconciliation procedures. The ratio of fresh data that needs to be transmitted, compared to old data that is already available at both endpoints of a reconciliation procedure, decreases with the lifetime of such a system. The overall size of the local set held by a node will eclipse the size of the symmetric difference in many reconciliation sessions.

There exist sophisticated protocols~\cite{eppstein2011s} that solve set reconciliation in a constant number of communication rounds and with communication complexity linear in the size of the symmetric difference. These impressive bounds are provably optimal~\cite{minsky2003set}, but computing the necessary messages requires time and space linear in the full size of the set held by a particular node, even if the symmetric difference of the two sets to reconcile is small.

These computational costs can be prohibitive, particularly in a peer-to-peer setting where frequent, optimistic reconciliations attempts are required to propagate information quickly and thoroughly. We hence look for an alternative that sacrifices optimal communication complexity for a bound on the computational complexity based on the size of the symmetric difference. See~\cite{auvolat2019merkle} for a publication with a greater focus on the anty-entropy use-case of set reconciliation, we focus primarily on the reconciliation technique itself.

Range-based set reconciliation is an approach in which nodes can perform the necessary computations within space proportional to the size of the symmetric difference of the two sets, and in time proportional to the size of the symmetric difference or the logarithm of the size of the local set, whichever is greater. This comes at the cost of a logarithmic number of communication rounds, as the procedure follows a straightforward divide-and-conquer approach: the sets are sorted according to some total order, and nodes initiate reconciliation by sending the fingerprint of all their local items within a certain range. Upon receiving a pair of range delimiters and a fingerprint, a node computes the fingerprint of all of its own local items within that range. If the fingerprints match, the range has been successfully reconciled. Otherwise the node splits the range into subranges, such that it has roughly the same number of items in each subrange, and initiates reconciliation of these new ranges. Whenever a node receives the fingerprint of the empty set, it transmits all its local items within that range to its peer.

If the fingerprint of a set can be computed by associatively combining the fingerprints of its members, for example, by using the exclusive or of hashes of all members, then we can compute it efficiently: we store the set as a balanced search tree, in which every vertex also stores the fingerprint of its subtree. Maintaining these cached fingerprints takes no more time asymptotically than maintaining the tree structure itself, and we can compute the fingerprint for any subrange by traversing the tree in logarithmic time.

This approach appears as an implementation detail for a specific use case in~\cite{chen1999prototype} Section 3.6 and~\cite{shang2017survey} Section II.A\footnote{We cite a survey because there is no standalone publication on CCNx 0.8 Sync. The survey refers to online documentation at \url{https://github.com/ProjectCCNx/ccnx/blob/master/doc/technical/SynchronizationProtocol.txt}}, but neither publication studies it as a viable solution to the set reconciliation problem in its own right. To the best of our knowledge, there is no literature specifically dedicated to the range-based approach. We aim to mitigate this with a comprehensive overview, hopefully limiting further reinvention of the wheel.

% A thorough inspection of the data structures involved in range-based set reconciliation leads to interesting parallels to the design of authenticated set data structures~\cite{naor2000certificate}. In investigating cryptographically secure choices for reconciliation, we create authenticated data structures that do not prescribe specific tree shapes in order to compute digests and certificates.

Beyond a more general and precise formulation and a more detailed complexity analysis than in~\cite{chen1999prototype} and~\cite{shang2017survey}, we make several new contributions:

\begin{itemize}
    \item we reduce the time complexity of successive fingerprint computations by a logarithmic factor while using only a constant amount of space,
    \item we give an algebraic characterization of suitable fingerprint functions,
    \item we discuss cryptographically secure fingerprints and survey suitable candidates, and
    % \item we present the first authenticated set data structures that abstract over the exact tree representation of their set by using secure, associative hash functions, and
    \item we investigate the role that history-independent data structures can play for range-based set reconciliation.
\end{itemize}

The organization of this article is as follows: we first review related work in \cref{related-work}. We then state the protocol for range-based reconciliation in \cref{reconciliation} and analyze its complexity. In \cref{sec:computation} we examine possible choices of fingerprints and show how to compute them efficiently. In \cref{secure} we examine how malicious actors can influence reconciliation and survey secure fingerprint functions that can protect against this
%, which leads to our authenticated set data structures
. We then examine pseudorandom trees as a fingerprinting technique in \cref{randomization}, before concluding in \cref{conclusion}.

\section{Related Work}\label{related-work}

The range-based approach to set reconciliation has been employed as an implementation detail for a specific use case in~\cite{chen1999prototype} section 3.6 and~\cite{shang2017survey} section II.A. Neither work studies it as a self-contained approach of interest to the set reconciliation problem. Consequently the treatment is rather superficial, e.g., neither work considers collision resistance, even though it is relevant to both settings. Considering the simplicity of the approach, there may well be further publications reusing or reinventing it. To the best of our knowledge, there is however no prior work specifically dedicated to the approach.

We give an overview of the set reconciliation literature in \cref{related-reconciliation}, and of the existing literature on set fingerprinting in \cref{related-fingerprinting}.

\subsection{Set Reconcilliation}
\label{related-reconciliation}

Most reconciliation literature focuses on reconciliation in a single communication round, at the price of high computational costs. In the following presentations, we consider the setting of a node $\mathcal{X}_0$ holding a set $X_0 \subseteq U$ reconciling with a node $\mathcal{X}_1$ holding a set $X_1 \subseteq U$. We denote by $n_{\triangle}$ the size of the symmetric difference between $X_0$ and $X_1$.

The seminal work on set reconciliation is~\cite{minsky2003set}, introducing an approach based on characteristic polynomial interpolation (CPI). If an approximation of the size $n_{\triangle}$ of the symmetric difference of the two sets to be reconciled is known, the total number of bits that need to be transmitted is proportional to $n_{\triangle}$, which is more efficient than the range-based approach. Nodes need to interpolate certain polynomials however, which the authors reduced to performing Gaussian elimination, which takes $\complexity{n_{\triangle}^3}$ time.

The authors further propose a strategy for determining an approximation of $n_{\triangle}$ over a logarithmic number of communication rounds. When using this strategy, CPI requires the same number of roundtrips as range-based reconciliation, but at higher computational complexity.

A bloom filter~\cite{bloom1970space} is a probabilistic data structure for set membership queries. Invertible bloom lookup tables~\cite{goodrich2011invertible} (IBLTs) extend this behavior with the ability to list all items stored in the data structure, albeit again with an error probability which decreases as the fraction of space per contained item goes up. \cite{eppstein2011s}~introduces set reconciliation based on IBLTs by formulating a way to compute the difference between two IBLTs, with the result corresponding to an IBLT encoding the corresponding set difference. Creating the required IBLT requires $\complexity{\abs{X_i}}$ time for node $\mathcal{X}_i$ and $\complexity{n_{\triangle}}$ space, both of which dominate the cost for subtracting and inverting them.

Similar to the CPI approach, the nodes need an estimate of the size of the set difference before they can perform reconciliation. The authors present a single-message estimation protocol for the set difference that uses IBLTs as well. The size of the message is in $\complexity{\log(\abs{U})}$. Both creating or receiving and using the message requires $\complexity{X_i}$ time for node $\mathcal{X}_i$ and $\complexity{\log(\abs{U})}$ space.

Overall, the IBLT approach achieves set reconciliation in a single round trip, transmitting only $\complexity{n_{\triangle} + \log(\abs{U})}$ bits. The computational cost is however linear in the size of the sets, and the space requirements for the computation are in $\complexity{n_{\triangle}}$, which can cause trouble in the case of lopsided reconciliation sessions:

Several other approaches based on bloom filters have been published, all of which achieve a constant number of roundtrips and a small message size at the cost of at least linear computation time and computation space requirements.

\cite{byers2002fast}~transmits the nodes of a patricia tree representation of the sets in a bloom filter, \cite{tian2011exact} uses a bloom filter-based approach to obtain an estimate of the size of the set difference in the first phase of a CPI reconciliation. \cite{guo2012set} employs counting bloom filters, \cite{luo2019set} makes use of cuckoo filters. \cite{ozisik2019graphene}~combines IPLTs with regular bloom filters to reduce the message size.

\cite{minsky2002practical} introduces \defined{partition reconciliation},which reconciles in a logarithmic number of rounds in order to reduce computational load. It builds upon the CPI approach, and attempts reconciliation for successively smaller subsets until it succeeds once the subsets becomes small enough.

This approach eliminates CPI's cubic scaling of the computation time in $n_{\triangle}$. Similar to our auxiliary tree, the authors propose a tree structure of partitions where a parent node includes subranges as children. The reconciliation message for each of these ranges is precomputed and stored within the tree.

Given such a partition tree, the reconciliation procedure has the same asymptotic worst-case complexity bounds as ours. It has performs better on average however: whereas our approach recurses whenever the two ranges that are being compared are unequal, i.e., whenever the size of the symmetric difference is greater than zero, their approach only recurses whenever the size of the symmetric difference is greater than some arbitrary, fixed constant.

The tree can be updated over insertions and deletions in time logarithmic in the height of the tree, but these updates are not balancing. In the worst case, maintaining this auxiliary data structure can thus take time linear in the size of the set per update. Furthermore, the tree is specific to a particular choice of evaluation points and control points for the characteristic polynomial. If the tree is to be reused across multiple reconciliation sessions, these points have to be fixed in advance. This could allow an attacker to craft malicious sets for which a failed reconciliation is not detected. As a consequence, this approach can only be used with the producers of the sets are trusted.\

\cite{auvolat2019merkle} also considers reconciliation in a logarithmic number of rounds, they use pseudorandom Merkle trees, comparable to our \cref{randomization}. Their approach is far less flexible than ours however: reconciliation messages are guided by the tree shape, so the number of recursion steps in each round is fixed, and the maximum number of rounds can degrade to $\complexity{n}$ in the worst case. Our approach leaves complete freedom for the number of recursion steps and guarantees a logarithmic number of communication rounds in the worst case, even when using a pseudorandom tree for fingerprint computations. Their promising experimental evaluation provides a strong indicator that our more efficient approach would also perform well in practice.

\subsection{Fingerprinting Sets}
\label{related-fingerprinting}

As range-based synchronization relies on fingerprinting sets, we give a small overview of work on the topic.

Merkle trees~\cite{merkle1989certified} introduced the idea of maintaining hashes in a tree to efficiently compute and update a fingerprint. Since the exact shape of the tree determines the root label, unique representations of sets as trees are of interest. \cite{uniquerepresentation} proves the important negative result that unique representations require superlogarithmic time to maintain under insertions and deletions in general. \cite{sundar1994unique} gives logarithmic solutions for sparse or dense sets and points to further deterministic solutions.

Unique representations with probabilistic time complexities that are logarithmic in the expected case have been studied in~\cite{pugh1989incremental}, suggesting hash-tries as a set representation for computing hashes. Pugh also developed skip lists~\cite{pugh1990skip}, a probabilistic set representation not based on trees. \cite{seidel1996randomized} offers treaps, another randomized tree structure. Further study of uniquely represented data structures has been done under the moniker of \textit{history-independent data structures}, including IO-efficient solutions for treaps~\cite{golovin2009b} and skip list~\cite{bender2016anti} in an \textit{external memory} computational model. Using history-independent data structures for defining the fingerprints in a range-based reconciliation scheme is possible in principle, but malicious actors can craft degenerate sets that incur super-logarithmic computation times. Furthermore, such an approach cannot leave the exact choice of set data structure as an implementation detail.

Beyond the comparison of root labels for set equality testing, Merkle trees and their technique of hashing the concatenation of hashes form the basis of many authenticated data structures, ranging from simple balanced trees or treaps~\cite{naor2000certificate}, authenticated maps~\cite{buldas2000accountable}\cite{auvolat2019merkle} and skip lists~\cite{goodrich2000efficient} to more general DAG-based data structures~\cite{martel2004general}.

A different line of authenticated data structures utilizes dynamic accumulators~\cite{camenisch2002dynamic}, small digests for sets that can be efficiently updated under insertion and deletion of items, and which allow computing small certificates that some item has been ingested. \cite{papamanthou2016authenticated} and \cite{papamanthou2011optimal} use accumulators to provide authenticated set data structures that are more efficient than their Merkle-tree-based counterparts. Accumulators are stronger than necessary for the range-based synchronization approach, so they are not discussed in our main text.

Orthogonal to these lines of research are algebraic approaches. The earliest mention of hashing into a group and performing computations on hashes to compactly store information about sets that we could find is in~\cite{wegman1981new}. They provide probabilistic set equality testing, but without maintaining any tree structure, effectively anticipating (cryptographically insecure) homomorphic set fingerprinting.

The first investigation of the cryptographic security of this approach was done in the seminal~\cite{bellare1997new}, albeit in the slightly less natural context of sequences rather than sets, with the additive hash being broken in~\cite{wagner2002generalized} and~\cite{lyubashevsky2005parity}. Multiset homomorphic terminology is due to~\cite{clarke2003incremental}; \cite{cathalo2009comparing},~\cite{maitin2017elliptic} and~\cite{lewi2019securing} give further constructions.

The generalized hash tree of~\cite{papamanthou2013streaming} extends the idea of combining hashes via a binary operation to operations that are not closed, instead the output is mapped back into the original domain by a separate function. These ``compressed'' outputs are then arranged in a tree; this technique allows using the algebraic approach for authenticated data structures.

\section{Recursive Reconciliation}\label{reconciliation}

In this section, we describe the communication that goes into a range-based set reconciliation session, while deferring the details of fingerprint computations to \cref{sec:computation}. 

We consider the setting of two nodes \peer{0} and \peer{1} that are connected via a bidirectional, reliable, ordered communication channel. We assume that an arbitrary number of bits can be sent in a single, unit-length communication round. The nodes initially hold sets $X_0$ and $X_1$ respectively, and after reconciliation, both will hold $X_0 \cup X_1$.

$X_0$ and $X_1$ are drawn from some universe $U$, which is ordered by a total order $\preceq$. To allow meaningful statements about communication complexity, we require every element from $U$ to be transmittable using a bounded number of bits, that is, we require $U$ to be finite. This can always be achieved in practice by reconciling hashes of items rather than items themselves.

We finally assume that there is a fingerprinting function $\fun{\fpname}{\powerset{U}}{H}$ that maps arbitrary subsets of $U$ (and hence also of  $X_0$ and $X_1$) into some finite codomain $H$ with negligible probability of collisions. We discuss such functions in detail in \cref{sec:computation}.

Before we can define the message exchange, we need some formal definitions and terminology for talking about ranges:

\begin{definition}[Range]
\label{def:ranges}
Let $S \subseteq U$  and $x, y \in U$.

The \defined{range from $x$ to $y$ in $S$}, denoted by $\range{x}{y}{S}$, is the set $\set{s \in S \mid x \preceq s \prec y}$ if $x \prec y$, or $S \setminus \range{y}{x}{S}$ if $y \prec x$, or simply $S$ if $x = y$. We call $x$ the \defined{lower boundary} and $y$ the \defined{upper boundary} of the range (even if $y \preceq x$).
\end{definition}

Note that $x$ and $y$ need not be elements of $S$ themselves.

\subsection{Protocol Description}

In a given communication round, a node receives information about some subranges of the sets to be reconciled: for each subrange in question, it receives either a fingerprint which depends exactly on which items the other node holds within that subrange, or it receives specific items within that subrange that need to be reconciled. The node then answers by supplying its own information, further partitioning ranges into arbitrarily many subranges if neither the received fingerprint matches the fingerprint of the local items within that range nor the range contains few enough items to transmit them directly. We precisely specify the vocabulary by which nodes exchange information in \cref{def:messages}:

\begin{definition}[Message]
\label{def:messages}
Let \peer{i} be a node that holds a set $X_i \subseteq U$.

A \defined{range fingerprint} is a triplet $\ifp{x}{y}{X_i}$ for $x, y \in U$. It conveys the fingerprint over the range from $x$ to $y$ in $X_i$.

A \defined{range item set} is a four-tuple $\iis{x}{y}{S}{b}$ for $x, y \in U$, $S \subseteq \range{x}{y}{X_i}$, and $b \in \{0, 1\}$. It transmits items within the range from $x$ to $y$ in $X_i$. The boolean flag signals whether the other node should respond with its local items from $x$ to $y$ as well ($b = 0$), or whether these have already been received ($b = 1$).
 
A \defined{message part} is either a range fingerprint or a range item set. A \defined{message} is a nonempty sequence of message parts. 
\end{definition}

\theoremstyle{definition}
\newtheorem{protocol}{Protocol}
\Crefformat{protocol}{#2Protocol~#1#3}
\crefformat{protocol}{#2protocol~#1#3}
\Crefname{protocol}{Protocol}{Protocols}
\crefname{protocol}{protocol}{protocols}

In order to initiate reconciliation, a node sends a message consisting solely of a range fingerprint $\ifp{x}{x}{X_i}$ for some $x \in U$. The nodes can then reconcile their sets by following \cref{protocol:rbsr}:

\begin{protocol}[Range-Based Set Reconciliation]
\label{protocol:rbsr}

Let \peer{i} be a node that holds a set $X_i \subseteq U$ and that has just received a message. It then performs the following actions:

\begin{enumerate}
    \item[1)] Initialize an empty response.
    \item[2)] For every range item set $\iis{x}{y}{S}{b}$ in the message, add $S$ to $X_i$. If $b = 0$ and $\range{x}{y}{X_i} \setminus S \neq \emptyset$, add the range item set $\iis{x}{y}{\range{x}{y}{X_i} \setminus S}{1}$ to the response.
    \item[3)] For every range fingerprint $\ifp{x}{y}{X_j}$ in the message, do one of the following:
        \begin{enumerate}
            \item[] \textbf{Case 1, Equal Fingerprints:}\newline
			If $\fp{\range{x}{y}{X_j}} = \fp{\range{x}{y}{X_i}}$, do nothing.
            \item[] \textbf{Case 2, Recursion Anchor:} You may add the range items set $\iisnatural{x}{y}{X_i}{0}$ to the response.\newline
			If $\abs{\range{x}{y}{X_i}} \leq 1$ or $\fp{\range{x}{y}{X_j}} = \fp{\emptyset}$, you must do so.
            \item[] \textbf{Case 3, Recurse:} Otherwise, select $m_0 \defeq x \prec m_1 \prec \ldots \prec m_k \defeq y$ from $U$, $k \geq 2$, such that among all $\range{m_l}{m_{l + 1}}{X_i}$ for  $0 \leq l < k$ at least two ranges are nonempty. For all $0 \leq l < k$ add either the range fingerprint $\ifp{m_l}{m_{l + 1}}{X_i}$ or the range item set $\iisnatural{m_l}{m_{l + 1}}{X_i}{0}$ to the response.
        \end{enumerate}
    \item[4)] If the accumulated response is nonempty, send it. Otherwise terminate successfully.
\end{enumerate}
\end{protocol}

\Cref{simple-set-reconciliation-example} visualizes an example run of the protocol.

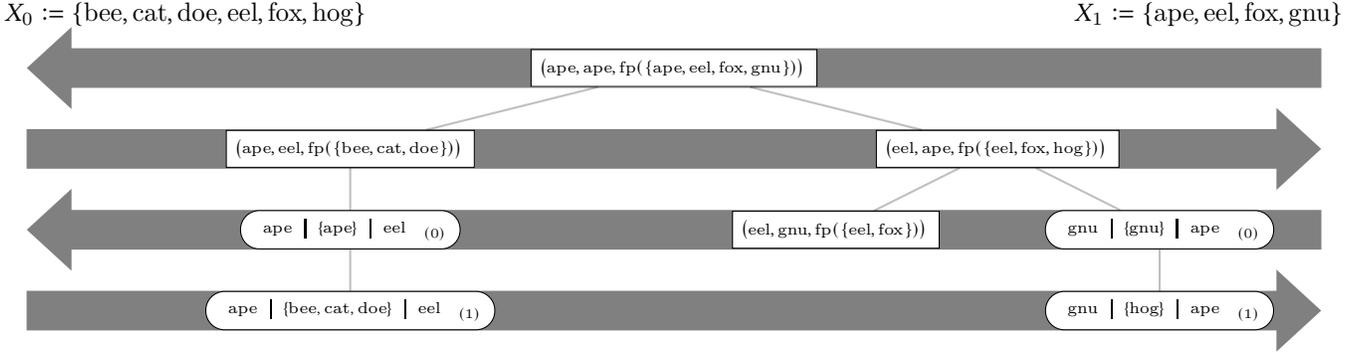
\begin{figure*}
$X_0 \defeq \{\exampleb, \examplec, \exampled, \examplee, \examplef, \exampleh \}$
\hfill
$X_1 \defeq \{\examplea, \examplee, \examplef, \exampleg\}$

\begin{scaletikzpicturetowidth}{\textwidth}
\begin{tikzpicture}[scale=\tikzscale, font=\tiny]
	\pgfdeclarelayer{background}
	\pgfdeclarelayer{foreground}
	\pgfsetlayers{background,main,foreground}

	\begin{pgfonlayer}{main}
		%vertices
		\node (vroot) at (0, 1) [fpi] {\examplefpi{\examplea}{\examplei}{\{\examplea, \examplee, \examplef, \exampleg\}}};

		\node (v00) at (-4, -0) [fpi] {\examplefpi{\examplea}{\examplee}{\{\exampleb, \examplec, \exampled\}}};
		\node (v01) at (4, -0) [fpi] {\examplefpi{\examplee}{\examplei}{\{\examplee, \examplef, \exampleh\}}};

                \node (v10) at (-4, -1) [iis] {\exampleiis{\examplea}{\examplee}{\{\examplea\}}{0}};
                \node (v11) at (2, -1) [fpi] {\examplefpi{\examplee}{\exampleg}{\{\examplee, \examplef\}}};
                \node (v12) at (6, -1) [iis] {\exampleiis{\exampleg}{\examplei}{\{\exampleg\}}{0}};

                \node (v20) at (-4, -2) [iis] {\exampleiis{\examplea}{\examplee}{\{\exampleb, \examplec, \exampled\}}{1}};
                \node (v21) at (6, -2) [iis] {\exampleiis{\exampleg}{\examplei}{\{\exampleh\}}{1}};
		%edges
                \draw (vroot) edge[edge] (v00);
                \draw (vroot) edge[edge] (v01);

		\draw (v00) edge[edge] (v10);
		\draw (v01) edge[edge] (v11);
		\draw (v01) edge[edge] (v12);

		\draw (v10) edge[edge] (v20);
		\draw (v12) edge[edge] (v21);
	\end{pgfonlayer}

	\begin{pgfonlayer}{background}
		\draw[-{Triangle[width=30pt,length=17pt,color=gray]}, line width=15pt, color=gray](8, 1) -- (-8, 1);
		\draw[-{Triangle[width=30pt,length=17pt,color=gray]}, line width=15pt, color=gray](-8, -0) -- (8, -0);
		\draw[-{Triangle[width=30pt,length=17pt,color=gray]}, line width=15pt, color=gray](8, -1) -- (-8, -1);
		\draw[-{Triangle[width=30pt,length=17pt,color=gray]}, line width=15pt, color=gray](-8, -2) -- (8, -2);
	\end{pgfonlayer}
\end{tikzpicture}
\end{scaletikzpicturetowidth}

\caption[Detailed synchronization example]{
An example run of \cref{protocol:rbsr}. In this and further examples, $U \defeq \set{\examplea, \exampleb, \examplec, \exampled, \examplee, \examplef, \exampleg, \exampleh}$, and $\preceq$ orders the universe alphabetically. Range fingerprints have sharp corners, range item sets have rounded corners. The arrows in the background indicate which node is sending and which node is receiving.

\peer{1} initiates reconciliation over the full universe, transmitting the fingerprint of $X_1$.

Upon receiving this range fingerprint, \peer{0} locally computes $\fp{\range{\examplea}{\examplei}{X_0}}$. Because the result does not match the received fingerprint, \peer{0} splits $X_0$ into two parts of equal size and transmits range fingerprints for these subranges.

In the third round, \peer{1} locally computes fingerprints for the two received ranges, but neither matches. Because $\abs{\range{\examplea}{\examplee}{X_1}} \leq 1$, \peer{1} transmits the corresponding range items set  $\iis{\examplea}{\examplee}{\examplea}{0}$. For the other range, we have that $\abs{\range{\examplee}{\examplei}{X_1}} > 1$ however, so another recursion step can be performed. After splitting the range, the lower range contains enough items to send another range fingerprint. The upper range however only contains one item, thus \peer{1} handles it by sending a range item set.

In the fourth and final communication round, \peer{0} receives the two range item sets and answers with the items it holds within those ranges. For the range fingerprint $\ifp{\examplee}{\exampleg}{X_1}$, it computes an equal fingerprint $\fp{\range{\examplee}{\exampleg}{X_0}}$, so that range can be considered successfully reconciled without transmitting any more data.
}

\label{simple-set-reconciliation-example}
\end{figure*}

\subsection{Protocol Properties}

\Cref{protocol:rbsr} gives some leeway for nodes to decide whether and into how many subranges to split any range fingerprint they receive. The precise choices do not impact the correctness of the protocol. In particular, two nodes can reconcile a set even when using different strategies for deciding when and how to recurse. In a decentralized setting, different implementations do not need to coordinate and can make decisions based on their available resources.

\subsubsection{Termination and Correctness}

Termination of \cref{protocol:rbsr} follow from an inductive argument. The key observation is that the number of items in the largest subrange of the current message is strictly decreasing with each round (even exponentially so). Small ranges and ranges with matching fingerprints are handled within a constant number of communication rounds, so the protocol always terminates, even if the nodes prefer recursing over sending range item whenever allowed to do so.

Correctness can also be argued inductively. Under the assumption that fingerprints do not collide, ranges with equal fingerprints are reconciled correctly. Sending a range item set and receiving the response also leads to both nodes storing the union of all items within that range. The subranges created by the recursive case are reconciled correctly by induction hypothesis. And because the subranges partition the original range\footnote{In fact, for correctness it already suffices that the subranges \textit{cover} the original range. We limit our discussion to \textit{partitions} because they minimally cover the original range without containing any duplicate items.}, this completely reconciles the original range.

\subsubsection{Complexity}

The protocol gives nodes the freedom to respond to any range fingerprint with a range item set, even if the range fingerprint is arbitrarily large. For a meaningful complexity analysis we need to restrict the behavior of nodes. A realistic modus operandi is for a node to send a range item set only if the number of its items in the range is less than or equal to some threshold $t \in \Np$. Higher choices for $t$ reduce the number of roundtrips, but increase the probability that an item is being sent even though the other node already holds it.

A node is similarly given freedom over the number of subranges into which to split a range when recursing. We will assume that nodes always split into at most $b \in \N, b \geq 2$ subranges. As with $t$, higher numbers reduce the number of roundtrips, at the cost of potentially sending fingerprints that did not need sending.

Transmitting all available items within a range is a very simple choice of recursion anchor, but becomes wasteful for large choices of $t$. A more sophisticated approach is to run a more efficient constant-round set reconciliation protocol such as~\cite{minsky2003set} once a range contains few enough items for both nodes. As that protocol would only be run on small sets, its time and space complexities would be in $\complexity{1}$, so the overall time and space complexities would be the same as for unmodified range-based reconciliation. Hence, range-based set reconciliation can be regarded as a mechanism for reducing the computational complexity of arbitrary reconciliation protocols. In our further analysis, we stick to \cref{protocol:rbsr} as presented however.

Because we want to analyze not only the worst-case complexity but also the complexity depending on the similarity between the two sets held by the participating nodes, we define some fine-grained instance size parameters: $n_0$ and $n_1$ denote the number of items held by \peer{0} and \peer{1} respectively. We let $n \defeq n_0 + n_1$, $n_{\min} \defeq \min(n_0, n_1)$ and $n_{\triangle} \defeq \abs{(X_0 \cup X_1) \setminus (X_0 \cap X_1}$ (the size of the symmetric difference of $X_0$ and $X_1$).

A helpful observation for the coming analyses is that the range fingerprints of a protocol run form a rooted, $b$-ary tree, compare \cref{simple-set-reconciliation-example}. When a leaf of the tree is reached, an exchange of range item sets follows.

Node $\mathcal{X}_i$ can perform at most $\ceil{\log_{b}(n_i)}$ recursion steps, so the overall height of the tree is bounded by $2 \cdot\ceil{\log_{b}(n_{\min})}$. The number of vertices of a $b$-ary tree of height $h$ is less than twice the number of leaves, so the overall number of vertices is in $\complexity{n}$.

% Node $\mathcal{X}_i$ can perform at most $\ceil{\log_{b}(n_i)}$ recursion steps, so the overall height of the tree is bounded by $2 \cdot\ceil{\log_{b}(n_{\min})}$. The number of vertices of a $b$-ary tree of height $h$ is at most $\sum_{i=0}^{h} b^{i} = \frac{b^{h} - 1}{b - 1}$. For $h \leq 2 \cdot\ceil{\log_{b}(n_{\min})}$, we have $$\frac{b^{h} - 1}{b - 1} \leq 2 \cdot 2 \cdot n_{\min} \leq 2n \in \complexity{n}.$$

The parameter $t$ determines when recursion is cut off, and thus influences the height of the tree. For $t = 1$, the protocol recurses as far as possible. For $t = b$, the last level of recursion is cut off, for $t = b^2$ the last two levels, and so on. Overall, the height of the tree is reduced by $\floor{\log_{b}(t)}$.

The total number of communication rounds required for reconciliation is bounded by the number of times that each node can split ranges without transmitting items, followed by two rounds of exchanging range items sets. This corresponds to two plus the height of the tree, so $2 + 2 \cdot\ceil{\log_{b}(n_{\min})} - \floor{\log_{b}(t)} \in \complexity{\log(n)}$.

This number cannot be bounded by $n_{\triangle}$, as witnessed by problem instances where one node is missing exactly one item compared to the other node. In such an instance, $b - 1$ branches in each recursion step result in equal fingerprints, but the one branch that does continue reaches the recursion anchor only after the full number of rounds. See \cref{fig:worst-rounds} for an example.

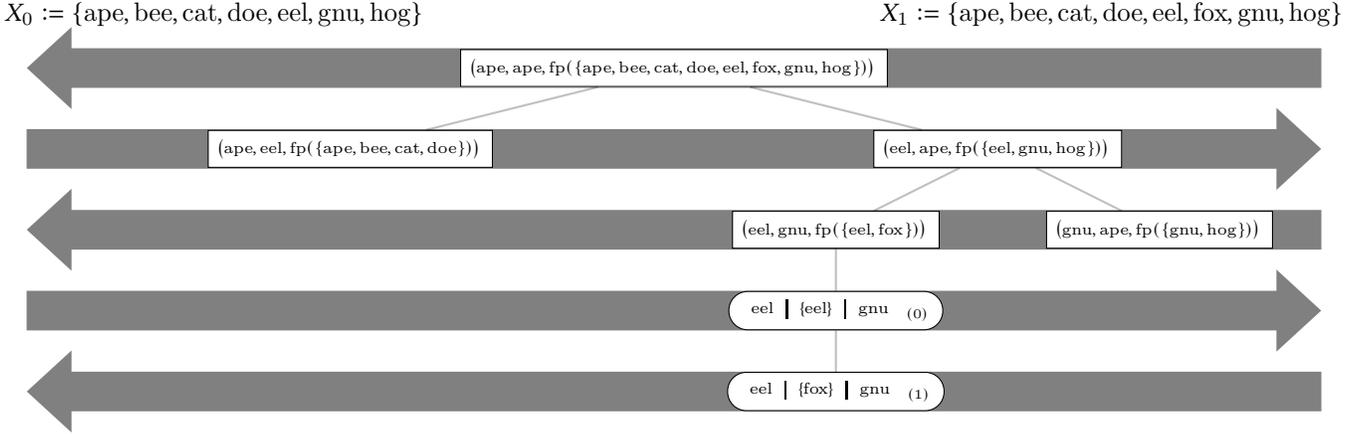
\begin{figure*}
$X_0 \defeq \{\examplea, \exampleb, \examplec, \exampled, \examplee, \exampleg, \exampleh \}$
\hfill
$X_1 \defeq \{\examplea, \exampleb, \examplec, \exampled, \examplee, \examplef, \exampleg, \exampleh \}$

\begin{scaletikzpicturetowidth}{\textwidth}
\begin{tikzpicture}[scale=\tikzscale, font=\tiny]
	\pgfdeclarelayer{background}
	\pgfdeclarelayer{foreground}
	\pgfsetlayers{background,main,foreground}

	\begin{pgfonlayer}{main}
		%vertices
		\node (vroot) at (0, 1) [fpi] {\examplefpi{\examplea}{\examplei}{\{\examplea, \exampleb, \examplec, \exampled, \examplee, \examplef, \exampleg, \exampleh\}}};

		\node (v00) at (-4, -0) [fpi] {\examplefpi{\examplea}{\examplee}{\{\examplea, \exampleb, \examplec, \exampled\}}};
		\node (v01) at (4, -0) [fpi] {\examplefpi{\examplee}{\examplei}{\{\examplee, \exampleg, \exampleh\}}};

                \node (v10) at (2, -1) [fpi] {\examplefpi{\examplee}{\exampleg}{\{\examplee, \examplef\}}};
                \node (v11) at (6, -1) [fpi] {\examplefpi{\exampleg}{\examplei}{\{\exampleg, \exampleh\}}};

                \node (v20) at (2, -2) [iis] {\exampleiis{\examplee}{\exampleg}{\{\examplee\}}{0}};

                \node (v30) at (2, -3) [iis] {\exampleiis{\examplee}{\exampleg}{\{\examplef\}}{1}};

		%edges
                \draw (vroot) edge[edge] (v00);
                \draw (vroot) edge[edge] (v01);

		\draw (v01) edge[edge] (v10);
		\draw (v01) edge[edge] (v11);

		\draw (v10) edge[edge] (v20);

		\draw (v20) edge[edge] (v30);
	\end{pgfonlayer}

	\begin{pgfonlayer}{background}
		\draw[-{Triangle[width=30pt,length=17pt,color=gray]}, line width=15pt, color=gray](8, 1) -- (-8, 1);
		\draw[-{Triangle[width=30pt,length=17pt,color=gray]}, line width=15pt, color=gray](-8, -0) -- (8, -0);
		\draw[-{Triangle[width=30pt,length=17pt,color=gray]}, line width=15pt, color=gray](8, -1) -- (-8, -1);
		\draw[-{Triangle[width=30pt,length=17pt,color=gray]}, line width=15pt, color=gray](-8, -2) -- (8, -2);
		\draw[-{Triangle[width=30pt,length=17pt,color=gray]}, line width=15pt, color=gray](8, -3) -- (-8, -3);
	\end{pgfonlayer}
\end{tikzpicture}
\end{scaletikzpicturetowidth}

\caption{An example run of the protocol that takes the greatest possible number of rounds, even though $n_{\triangle} = 1$. $b \defeq 2, t \defeq 1$.}

\label{fig:worst-rounds}

\end{figure*}

Every item in the symmetric difference can be responsible for only one such a path from the root to a leaf, a fact we can use to bound the number of transmitted bits. Range fingerprints and range item sets can be encoded using $\complexity{1}$ bits (because we assume $U$ to be finite and limit the size of range item sets to $t$). As the height of the tree is in $\complexity{\log(n)}$, we get an overall bound  of $\complexity{n_{\triangle} \cdot \log(n)}$ bits.

These paths overlap however, and every vertex of the tree only contributes $\complexity{1}$ bits to the reconciliation session, no matter how many items made it necessary to transmit that fingerprint. As we have at most $\complexity{n}$ nodes in the tree, the overall number of bits is at most $\complexity{\min(n_{\triangle} \cdot \log(n), n)}$. The more overlap between the paths, the fewer bits per item in the symmetric difference have to be transmitted. The case of transmitting $\complexity{n}$ bits occurs if one node is lacking every second item of the other node, see \cref{fig:worst-bytes}.

In terms of bits per item, this is very efficient however: in this scenario, $n_{\triangle}$ is within a constant factor of $n$, so we transmit $\complexity{1}$ bits per item that needs synchronization. The least efficient scenario from this point of view is that of $n_{\triangle} = 1$, where we send $\complexity{\log(n)}$ bits per item.

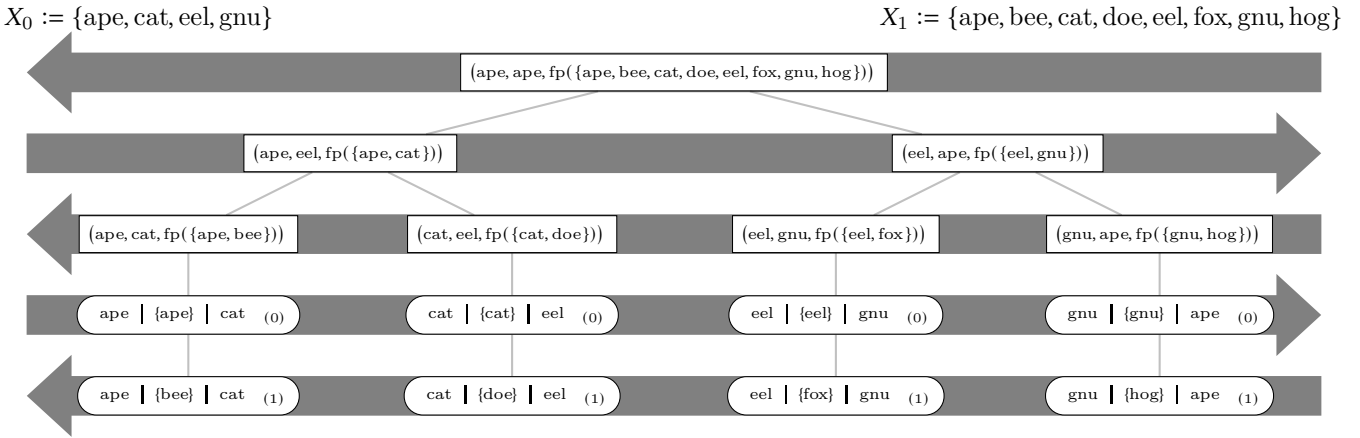
\begin{figure*}
$X_0 \defeq \{\examplea, \examplec, \examplee, \exampleg \}$
\hfill
$X_1 \defeq \{\examplea, \exampleb, \examplec, \exampled, \examplee, \examplef, \exampleg, \exampleh\}$

\begin{scaletikzpicturetowidth}{\textwidth}
\begin{tikzpicture}[scale=\tikzscale, font=\tiny]
	\pgfdeclarelayer{background}
	\pgfdeclarelayer{foreground}
	\pgfsetlayers{background,main,foreground}

	\begin{pgfonlayer}{main}
		%vertices
		\node (vroot) at (0, 1) [fpi] {\examplefpi{\examplea}{\examplei}{\{\examplea, \exampleb, \examplec, \exampled, \examplee, \examplef, \exampleg, \exampleh\}}};

		\node (v00) at (-4, -0) [fpi] {\examplefpi{\examplea}{\examplee}{\{\examplea, \examplec\}}};
		\node (v01) at (4, -0) [fpi] {\examplefpi{\examplee}{\examplei}{\{\examplee, \exampleg\}}};

                \node (v10) at (-6, -1) [fpi] {\examplefpi{\examplea}{\examplec}{\{\examplea, \exampleb\}}};
                \node (v11) at (-2, -1) [fpi] {\examplefpi{\examplec}{\examplee}{\{\examplec, \exampled\}}};
                \node (v12) at (2, -1) [fpi] {\examplefpi{\examplee}{\exampleg}{\{\examplee, \examplef\}}};
                \node (v13) at (6, -1) [fpi] {\examplefpi{\exampleg}{\examplei}{\{\exampleg, \exampleh\}}};

                \node (v20) at (-6, -2) [iis] {\exampleiis{\examplea}{\examplec}{\{\examplea\}}{0}};
                \node (v21) at (-2, -2) [iis] {\exampleiis{\examplec}{\examplee}{\{\examplec\}}{0}};
                \node (v22) at (2, -2) [iis] {\exampleiis{\examplee}{\exampleg}{\{\examplee\}}{0}};
                \node (v23) at (6, -2) [iis] {\exampleiis{\exampleg}{\examplei}{\{\exampleg\}}{0}};

                \node (v30) at (-6, -3) [iis] {\exampleiis{\examplea}{\examplec}{\{\exampleb\}}{1}};
                \node (v31) at (-2, -3) [iis] {\exampleiis{\examplec}{\examplee}{\{\exampled\}}{1}};
                \node (v32) at (2, -3) [iis] {\exampleiis{\examplee}{\exampleg}{\{\examplef\}}{1}};
                \node (v33) at (6, -3) [iis] {\exampleiis{\exampleg}{\examplei}{\{\exampleh\}}{1}};
		%edges
                \draw (vroot) edge[edge] (v00);
                \draw (vroot) edge[edge] (v01);

		\draw (v00) edge[edge] (v10);
		\draw (v00) edge[edge] (v11);
		\draw (v01) edge[edge] (v12);
		\draw (v01) edge[edge] (v13);

		\draw (v10) edge[edge] (v20);
		\draw (v11) edge[edge] (v21);
		\draw (v12) edge[edge] (v22);
		\draw (v13) edge[edge] (v23);

		\draw (v20) edge[edge] (v30);
		\draw (v21) edge[edge] (v31);
		\draw (v22) edge[edge] (v32);
		\draw (v23) edge[edge] (v33);
	\end{pgfonlayer}

	\begin{pgfonlayer}{background}
		\draw[-{Triangle[width=30pt,length=17pt,color=gray]}, line width=15pt, color=gray](8, 1) -- (-8, 1);
		\draw[-{Triangle[width=30pt,length=17pt,color=gray]}, line width=15pt, color=gray](-8, -0) -- (8, -0);
		\draw[-{Triangle[width=30pt,length=17pt,color=gray]}, line width=15pt, color=gray](8, -1) -- (-8, -1);
		\draw[-{Triangle[width=30pt,length=17pt,color=gray]}, line width=15pt, color=gray](-8, -2) -- (8, -2);
		\draw[-{Triangle[width=30pt,length=17pt,color=gray]}, line width=15pt, color=gray](8, -3) -- (-8, -3);
	\end{pgfonlayer}
\end{tikzpicture}
\end{scaletikzpicturetowidth}

\caption{An example run of the protocol that requires transmitting the maximum amount of bytes. $b \defeq 2, t \defeq 1$.}

\label{fig:worst-bytes}
\end{figure*}

Finally, we can observe that the largest possible size of a single message is proportional to the highest number of vertices in a single layer of the tree, i.e., it is proportional to $n_{\triangle}$. This affects the space complexity for the participating nodes. We will choose fingerprints such that nodes can process each message part within $\complexity{1}$ space. A simple way of implementing \cref{protocol:rbsr} is then for each node to store the full messages of size $\complexity{n_{\triangle}}$ in memory; the space requirement then also becomes $\complexity{n_{\triangle}}$.

Alternatively, nodes can split messages and transmit only a bounded number of message parts at a time, since it is not necessary to know the full message in order to process a single message part. Once a message fragment has been processed and the corresponding, newly computed response message parts have been sent, the other node can transmit the next message parts.

If both nodes follow this strategy and only allocate space for a constant number of message parts, the protocol can reach a deadlock. The number of message parts in a response can be greater than the number of message parts from which they were generated. When the number of message parts that would have to be transmitted exceeds the space capacity of both nodes added together, no node is able to receive or transmit more data.

It is however possible for only one of the two nodes to use a constant amount of space, whereas the other node then has to buffer up to a full outgoing and incoming message simultaneously, for a total space complexity of $\complexity{n_{\triangle}}$. This setup leads to a higher number of communication rounds. The node with unbounded memory has to wait for confirmation before transmitting a new set of message parts, which adds a round trip for every set of message parts that needs to be transmitted; transmitting a message of size $\complexity{n_{\triangle}}$ takes $\complexity{n_{\triangle}}$ communication rounds. The overall number of communication rounds for the protocol then becomes $\complexity{n_{\triangle} \cdot \log(n)}$.

This analysis seems unfavorable, but note that only a constant number of bits needs to be transmitted in every single communication round. The lower number of communication rounds in the setting where both nodes can store arbitrarily large messages can only be achieved if arbitrarily large messages can be transmitted in a single communication round. In a more realistic networking model, a limited bandwidth of $k$ bits per communication round implies that transmitting $n$ bits requires $\frac{n}{k} \in \complexity{n}$ rounds.

Under such a model, transmitting a single message of size $\complexity{n_{\triangle}}$ requires $\complexity{n_{\triangle}}$ rounds, hence, running range-based set reconciliation with unbounded memory takes $\complexity{n_{\triangle} \cdot \log(n)}$ rounds as well. So when taking finite bandwidth into account, one of the two nodes can operate with a constant amount of memory without affecting the asymptotic complexity of the protocol. Every possible set reconciliation protocol needs to transmit $\Omega(n_{\triangle})$ bits~\cite{minsky2003set}, so range-based set reconciliation is within a logarithmic factor of the optimal number of communication rounds under this model.

\section{Fingerprint Computation}\label{sec:computation}

We now examine the time and space complexity of the computations each node must perform during a reconciliation session. We consider a model where each node, in addition to working memory for performing computations, maintains an auxiliary data structure across computations. The node updates its auxiliary data structure whenever its set changes, and it can read from this data structure during its fingerprint computations.

This model is motivated by the fact that each node already has to update an external data structure --- its set  --- between message computations and in secondary storage rather than main memory. Overall, we are interested in the time it takes to update the auxiliary datastructure to reflect changes to the set, the space consumed by the auxiliary data structure, the time it takes to compute each message during a reconciliation session, and the space this requires.

Assuming the set is stored as a balanced search tree, it consumes a linear amount of space, and adding or removing individual items requires $\complexity{\log(n)}$ time. This gives us a free complexity budget to work with; if our auxiliary data structure requires the same amount of time and space, it does not impact the asymptotic performance of our approach. We will, in fact, extend the tree representation of the set by storing additional data in each vertex.

\subsection{Labeled Trees}

When computing messages, a node must be able to efficiently compute the fingerprint of all items it holds within arbitrary ranges. We now consider a general family of functions that map ranges within a set to some codomain, and that can be efficiently computed with an auxiliary tree structure. These functions reduce a finite set to a single value according to a monoid.

\begin{definition}[Monoid]
Let $M$ be a set, $\groupaddsym: M \times M \rightarrow M$, and $\neutraladd \in M$.

We call $(M, \groupaddsym, \neutraladd)$ a \defined{monoid} if it satisfies two properties:

  \begin{description}
    \item[associativity:] for all $x, y, z \in M$: $\groupadd{(\groupadd{x}{y})}{z} = \groupadd{x}{(\groupadd{y}{z})}$,
    \item[neutral element:] for all $x \in M$: $\groupadd{\neutraladd}{x} = x = \groupadd{x}{\neutraladd}$.
  \end{description}
\end{definition}

\begin{definition}[Lifted Function]
\label{def-lift}
Let $U$ be a set, $\preceq$ a linear order on $U$, $\mathcal{M} = (M, \groupaddsym, \neutraladd)$ a monoid, and $\fun{\f}{U}{M}$.

We \defined{lift $\f$ to finite sets via $\mathcal{M}$} to obtain $\partialfun{\lift{\f}{\mathcal{M}}}{\powerset{U}}{M}$ with:

\begin{align*}
\lift{\f}{\mathcal{M}}(\emptyset) &\defeq \mymathbb{0},\\
\lift{\f}{\mathcal{M}}(S) &\defeq \f\bigl(\min(S)\bigr) \oplus \lift{\f}{\mathcal{M}}\bigl(S \setminus \set{\min(S)}\bigr).\\
\end{align*}

In other words, if $S = \set{s_1, s_2, \ldots, s_{\abs{S}}}$ with $s_1 \prec s_2 \prec \cdots \prec s_{\abs{S}}$, then $\lift{\f}{\mathcal{M}}(S) = \groupadd{\f(s_1)}{\groupadd{\f(s_2)}{\groupadd{\cdots}{\f(s_{\abs{S}})}}}$.
\end{definition}

Let, for example, $U$ be an arbitrary set, $\mathcal{N}$ be the monoid of natural numbers under addition, and let $\lambda x.1$ map any $x$ to the number $1$, then $\lift{\lambda x.1}{\mathcal{N}}(S) = \abs{S}$ for every finite $S \subseteq U$.

We can efficiently compute lifted functions by maintaining a tree structure. We define trees inductively to simplify the presentation of algorithms:

\begin{definition}[Binary Tree]
A \defined{binary tree} $t$ over a universe $U$ is either the empty tree $\nil$, or a triplet of a left subtree $\tl{t}$, a value $\tv{t} \in U$, and a right subtree $\tr{t}$.

We say $t$ is a \defined{vertex} if $t \neq \nil$; we denote the set of all vertices in a tree $t$ by $\V(t)$. We say a vertex $t$ is a \defined{leaf} if $\tl{t} = \nil = \tr{t}$, otherwise, $t$ is \defined{internal}.

Let $\preceq$ be a total order. We say $t$ is a \defined{search tree} (with respect to $\preceq$) if $t = \nil$, or if $\tv{t}$ is greater than the greatest value in $\tl{t}$, $\tv{t}$ is less than the least value in $\tr{t}$, and every subtree of $t$ is also a search tree.
\end{definition}

Toward efficient computation of functions of the form $\lift{\f}{\mathcal{M}}$, we label a binary search tree $t$:

\begin{definition}[Monoid Tree]
Let $U$ be a set, $S \subset U$ a finite set, $\preceq$ a linear order on $U$, $\mathcal{M} \defeq (M, \groupaddsym, \neutraladd)$ a monoid, $\fun{\f}{U}{M}$, and let $t$ be a binary search tree on $S$.

We define a \defined{labeling function} $\fun{\liftlabel{\f}{\mathcal{M}}}{S}{M}$:\\
$\liftlabel{\f}{\mathcal{M}}(t) \defeq \neutraladd$ if $t = \nil$,\\
$\liftlabel{\f}{\mathcal{M}}(t) \defeq \groupadd{\liftlabel{\f}{\mathcal{M}}(\tl{t})}{\groupadd{\f(\tv{t})}{\liftlabel{\f}{\mathcal{M}}(\tr{t})}}$ otherwise.
We call a tree labeled by such a labeling function a \defined{monoid tree}.
\end{definition}

Observe that $\liftlabel{\f}{\mathcal{M}}(t) = \lift{\f}{\mathcal{M}}(\V(t))$ for every binary search tree $t$. The exact shape of the tree dictates the grouping of how to apply $\groupaddsym$ to several values; different groupings yields the same result, as $\groupaddsym$ is associative. Because we label a search tree, $\groupaddsym$ is always applied to the items in ascending order. If we were labeling arbitrary, not necessarily sorted binary trees, $\groupaddsym$ would have to be commutative for the equality of labels and lifted functions to hold.

Returning to our previous example, labeling a tree with $\liftlabel{\lambda x.1}{\mathcal{N}}$ annotates each subtree with its size, i.e., this yields the order statistic trees~\cite{cormen2022introduction}. The labels can be kept updated in a self-balancing search tree implementation without changing the asymptotic complexity of insertion and deletion for both $\liftlabel{\lambda x.1}{\mathcal{N}}$ in particular and for arbitrary $\liftlabel{\f}{\mathcal{M}}$ functions in general~\cite{cormen2022introduction}.

Every function of the form $\liftlabel{\f}{\mathcal{M}}$ for some function $\f$ and some monoid $\mathcal{M}$ can be used for maintaining labels in the tree, but do there exist other such functions as well? To answer this, we give a homomomorphism-flavored characterization of candidate functions: given the images of two sets, one containing only items strictly less than those in the other, the image of the union of these sets should be the same as combining the original images in some monoid. This ensures that vertex labels can be updated by considering only the labels of their children and the image of their value.

\begin{definition}[Tree-Friendly Function]
	Let $U$ be a set, $\preceq$ a linear order on $U$, $\mathcal{M} \defeq (M, \groupaddsym, \neutraladd)$ a monoid, and $\partialfun{\f}{\powerset{U}}{M}$ a partial function mapping all finite subsets of $U$ into $M$.
	
	We call $\f$ a \defined{\somewhatmorphism{}} if for all finite sets $S_0, S_1 \in \powerset{U}$ such that $\max(S_0) \prec \min(S_1)$, we have $\f(S_0 \cup S_1) = \groupadd{\f(S_0)}{\f(S_1)}$.
\end{definition}

This definition captures exactly the functions of form $\lift{\f}{\mathcal{M}}$, as shown in the following propositions:

\begin{proposition}
Let $U$ be a set, $\preceq$ a linear order on $U$, $\mathcal{M} \defeq (M, \groupaddsym, \neutraladd)$ a monoid, and $\fun{\f}{U}{M}$.

Then $\lift{\f}{\mathcal{M}}$ is a \somewhatmorphism{}.

\begin{proof}
Let $S_0, S_1 \in \powerset{U}$ be finite sets such that $\max(S_0) \prec \min(S_1)$. Then:

\begin{align*}
\lift{\f}{\mathcal{M}}(S_0 \cup S_1) &= \biggroupadd_{\substack{s_i \in S_0 \cup S_1,\\ \text{ascending}}} \f(s_i)\\
&= \biggroupadd_{\substack{s_i \in S_0,\\ \text{ascending}}} \f(s_i) \groupaddsym \biggroupadd_{\substack{s_i \in S_1,\\ \text{ascending}}} \f(s_i)\\
&= \lift{\f}{\mathcal{M}}(S_0) \groupaddsym \lift{\f}{\mathcal{M}}(S_1)\\
\end{align*}
\end{proof}
\end{proposition}

\begin{proposition}
Let $U$ be a set, $\preceq$ a linear order on $U$, $\mathcal{M} \defeq (M, \groupaddsym, \neutraladd)$ a monoid, and $\partialfun{\g}{\powerset{U}}{M}$ a \somewhatmorphism{}.

Then there exists $\fun{\f}{U}{M}$ such that $\g = \lift{\f}{\mathcal{M}}$.

\begin{proof}
Define $\fun{\f}{U}{M}$ as $\f(u) \defeq \g(\set{u})$. We show by induction on the size of $S \subseteq U$ that $\g(S) = \lift{f}{\mathcal{M}}(S)$.

\vspace{10pt}

\textbf{IB:} If $S = \emptyset$, then $\g(S) = \neutraladd = \lift{\f}{\mathcal{M}}(S)$. Suppose that $\g(\emptyset) \neq \neutraladd$, this would contradict the fact that for all $x \in U$ we have $\g(\set{x}) = \g(\set{x}) \groupaddsym \g(\emptyset) = \g(\emptyset) \groupaddsym \g(\set{x})$, which holds because $\set{x} = \set{x} \cup \emptyset = \emptyset \cup \set{x}$ and $\g$ is a \somewhatmorphism{}.

If $S = \set{x}$, then $\g(S) = \f(x) = \lift{\f}{\mathcal{M}}(S)$.

\textbf{IH:} For all sets $T$ with $\abs{T} = n$ it holds that $\g(T) = \lift{\f}{\mathcal{M}}(T)$.

\textbf{IS:} Let $S \subseteq U$ with $\abs{S} = n + 1$, then:

\begin{align*}
\g(S) &= \g(\set{\min(S)}) \groupaddsym \g(S \setminus \set{\min(S)})\\
&\overset{\text{IH}}= \g(\set{\min(S)}) \groupaddsym \lift{\f}{\mathcal{M}}(S \setminus \set{\min(S)})\\
&= \f(\min(S)) \groupaddsym \lift{\f}{\mathcal{M}}(S \setminus \set{\min(S)})\\
&= \lift{\f}{\mathcal{M}}(S)\\
\end{align*}

\vspace{10pt}
As $\g$ is only defined over finite inputs, we thus have $\g = \lift{\f}{\mathcal{M}}$.
\end{proof}
\end{proposition}

\subsection{Range Computations}

Given a monoid tree $t$, we can compute $\lift{\f}{\mathcal{M}}\bigl(\range{x}{y}{\V(t)}\bigr)$ efficiently for any $x, y \in U$. Without loss of generality, we can assume that $x \prec y$, as we can otherwise compute $\lift{\f}{\mathcal{M}}\bigl(\range{\min(\V(t))}{y}{\V(t)}\bigr) \groupaddsym \lift{\f}{\mathcal{M}}\bigl(\range{x}{\max(\V(t))}{\V(t)}\bigr)$.

Intuitively speaking, we trace paths from (the root of) $t$ to $x$ and $y$, and then we need to combine all values in the ``area between those paths''. The labels of the children of the vertices along these paths which lie within that area summarize this information, so it suffices to combine information that is available locally around the paths. If the tree is balanced, we thus only need to combine a logarithmic number of values.

\Cref{alg:aggregate_once} gives the precise definition of the algorithm. First, we search for the first vertex reachable from the root that lies within the range (the procedure \textproc{find\_initial}). If no such vertex exists, the set contains no items within the range. If such an initial vertex exists however, it is necessarily unique: assume toward a contradiction that there are two distinct such vertices $a \prec b$. Because $t$ is a search tree, the least common ancestor of $a$ and $b$ is also in the range, and it is closer to the root than both $a$ and $b$, a contradiction. Consequently, all items within the range are descendents of the initial vertex, which we name $init$.

Because all items within the range are descendents of $init$ and $x \preceq init \prec y$, we have that \begin{align*}
	\range{x}{y}{\V(t)} &= \set{z \in \V(\tl{init}) \mid z \succeq x} \disjointunion\\
	&\set{\tv{init}} \disjointunion \set{z \in \V(\tr{init}) \mid z \prec y},
\end{align*} and hence \begin{align*}
	\lift{\f}{\mathcal{M}}\bigl(\range{x}{y}{\V(t)}\bigr) &= \lift{\f}{\mathcal{M}}\bigl(\set{z \in \V(\tl{init}) \mid z \succeq x}\bigr) \groupaddsym\\
	&\f(\tv{init}) \groupaddsym \lift{\f}{\mathcal{M}}\bigl(\set{z \in \V(\tr{init}) \mid z \prec y}\bigr).
\end{align*}

Procedure \textproc{aggregate\_left} demonstrates how to compute $\lift{\f}{\mathcal{M}}\bigl(\set{z \in \V(\tl{init}) \mid z \succeq x}\bigr)$: starting from the initial vertex, we search for $x$, accumulating the labels of all right children, as well as the monoid values that correspond to those vertices on the search path that are greater than or equal to $x$. Similarly, procedure \textproc{aggregate\_right} computes $\lift{\f}{\mathcal{M}}\bigl(\set{z \in \V(\tr{init}) \mid z \prec y}\bigr)$. \Cref{fig:aggregate_once} depicts an example run.

\begin{algorithm}
\caption{Computing $\lift{\f}{\mathcal{M}}\bigl(\range{x}{y}{\V(t)}\bigr)$.}\label{alg:aggregate_once}
\begin{algorithmic}[1]
\Require $x \preceq y, t \neq \nil$
\Procedure{aggregate\_range}{$t, x, y$}
	\If{$x = y$}
		\State \return $\liftlabel{\f}{\mathcal{M}}(t)$
	\EndIf
	\State $init \gets \Call{find\_initial}{t, x, y}$
	\If{$init = \nil$}
		\State \return $\neutraladd$
	\Else
		\State $acc_l \gets \Call{aggregate\_left}{\tl{init}, x}$
		\State $acc_r \gets \Call{aggregate\_right}{\tr{init}, y}$
		\State \return $acc_l \groupaddsym \f(\tv{init}) \groupaddsym acc_r$
	\EndIf
\EndProcedure

\Procedure{find\_initial}{$t, x, y$}
	\While{$\true$}
		\If{$t = \nil$}
			\State \return $t$
		\ElsIf{$\tv{t} \prec x$}
			\State $t \gets \tr{t}$
		\ElsIf{$\tv{t} \succeq y$}
			\State $t \gets \tl{t}$
		\Else
			\State \return $t$
		\EndIf
	\EndWhile
\EndProcedure

\Procedure{aggregate\_left}{$t, x$}
	\State $acc \gets \neutraladd$
	\While{$\true$}
		\If{$t = \nil$}
			\State \return $acc$
		\ElsIf{$\tv{t} \prec x$}
			\State $t \gets \tr{t}$
		\ElsIf{$\tv{t} = x$}
			\State \return $\f(\tv{t}) \groupaddsym \liftlabel{\f}{\mathcal{M}}(\tr{t}) \groupaddsym acc$
		\Else
			\State $acc \gets \f(\tv{t}) \groupaddsym \liftlabel{\f}{\mathcal{M}}(\tr{t}) \groupaddsym acc$
			\State $t \gets \tl{t}$
		\EndIf
	\EndWhile
\EndProcedure

\Procedure{aggregate\_right}{$t, y$}
	\State $acc \gets \neutraladd$
	\While{$\true$}
		\If{$t = \nil$}
			\State \return $acc$
		\ElsIf{$\tv{t} \prec y$}
			\State $acc \gets acc \groupaddsym \liftlabel{\f}{\mathcal{M}}(\tr{t}) \groupaddsym \f(\tv{t})$
			\State $t \gets \tr{t}$
		\ElsIf{$\tv{t} = x$}
			\State \return $acc \groupaddsym \liftlabel{\f}{\mathcal{M}}(\tr{t}) \groupaddsym \f(\tv{t})$
		\Else
			\State $t \gets \tl{t}$
		\EndIf
	\EndWhile
\EndProcedure
\end{algorithmic}
\end{algorithm}

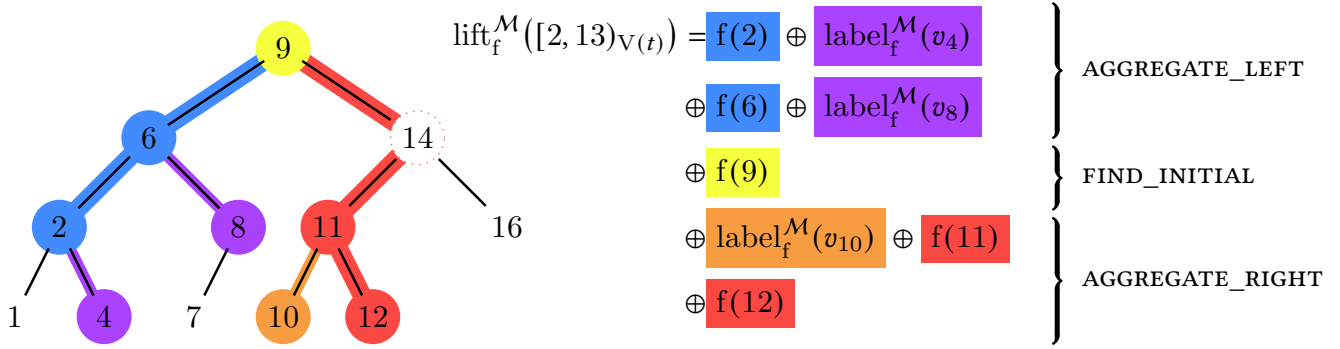
\begin{figure*}
\begin{adjustbox}{width=\textwidth}
\begin{tikzpicture}
\tikzstyle{vertex} = [opacity=1.0]
\tikzstyle{edge} = [draw,thick,opacity=1.0]

	\pgfdeclarelayer{background}
	\pgfdeclarelayer{foreground}
	\pgfsetlayers{background,main,foreground}
	
	\begin{pgfonlayer}{foreground}
		\node[vertex] (v1) at (1,1) {$1$};
		\node[vertex] (v2) at (1.5,2) {$2$};
		\node[vertex] (v4) at (2,1) {$4$};
		\node[vertex] (v6) at (2.5,3) {$6$};
		\node[vertex] (v7) at (3,1) {$7$};
		\node[vertex] (v8) at (3.5,2) {$8$};
		\node[vertex] (v9) at (4,4) {$9$};
		\node[vertex] (v10) at (4,1) {$10$};
		\node[vertex] (v11) at (4.5,2) {$11$};
		\node[vertex] (v12) at (5,1) {$12$};
		\node[vertex] (v14) at (5.5,3) {$14$};
		\node[vertex] (v16) at (6.5,2) {$16$};
		
		\draw (v2) edge[edge] (v1);
		\draw (v2) edge[edge] (v4);
		\draw (v6) edge[edge] (v2);
		\draw (v6) edge[edge] (v8);
		\draw (v8) edge[edge] (v7);
		\draw (v9) edge[edge] (v6);
		\draw (v9) edge[edge] (v14);
		\draw (v11) edge[edge] (v10);
		\draw (v11) edge[edge] (v12);
		\draw (v14) edge[edge] (v11);
		\draw (v14) edge[edge] (v16);
	\end{pgfonlayer}
	
	\begin{pgfonlayer}{background}
		\draw [leftSearch,line width=6pt] (v9.center) -- (v6.center) -- (v2.center);
		\draw [rightSearch,line width=6pt] (v9.center) -- (v14.center) -- (v11.center) -- (v12.center);
		\draw [leftInner,line width=3.5pt] (v6.center) -- (v8.center);				
		\draw [leftInner,line width=3.5pt] (v2.center) -- (v4.center);
		\draw [rightInner,line width=3.5pt] (v11.center) -- (v10.center);
		
		\filldraw [top] (v9.center) circle (0.3);
		\filldraw [leftSearch] (1.5,2) circle (0.3);
		\filldraw [leftSearch] (2.5, 3) circle (0.3);
		\filldraw [leftInner] (3.5,2) circle (0.3);
		\filldraw [leftInner] (2,1) circle (0.3);
		\filldraw [white] (v14.center) circle (0.3);
		\draw[rightSearch,dotted] (5.5,3) circle (0.3);
		\filldraw [rightSearch] (4.5,2) circle (0.3);
		\filldraw [rightSearch] (5,1) circle (0.3);
		\filldraw [rightInner] (4,1) circle (0.3);
	\end{pgfonlayer}

	\begin{pgfonlayer}{main}
	\node at (9,2.5) {
\begin{minipage}{6cm}
  \begin{align*}
    \lift{\f}{\mathcal{M}}\bigl(\range{2}{13}{\V(t)}\bigr) = &\colorbox{leftSearch}{$\f(2)$} \groupaddsym \colorbox{leftInner}{$\liftlabel{\f}{\mathcal{M}}(v_4)$} \\
    \groupaddsym& \colorbox{leftSearch}{$\f(6)$} \groupaddsym \colorbox{leftInner}{$\liftlabel{\f}{\mathcal{M}}(v_8)$} \\
    \groupaddsym &\colorbox{top}{$\f(9)$} \\
    \groupaddsym& \colorbox{rightInner}{$\liftlabel{\f}{\mathcal{M}}(v_{10})$} \groupaddsym \colorbox{rightSearch}{$\f(11)$} \\
    \groupaddsym& \colorbox{rightSearch}{$\f(12)$} \\
  \end{align*}
\end{minipage}
};

\draw [ultra thick, decorate, decoration = {calligraphic brace}] (12.6,4.5) --  (12.6,3.0);
\node[anchor=west] (c1) at (12.8,3.75) {\textproc{aggregate\_left}};

\draw [ultra thick, decorate, decoration = {calligraphic brace}] (12.6,2.9) --  (12.6,2.2);
\node[anchor=west] (c1) at (12.8,2.55) {\textproc{find\_initial}};

\draw [ultra thick, decorate, decoration = {calligraphic brace}] (12.6,2.1) --  (12.6,0.7);
\node[anchor=west] (c1) at (12.8,1.4) {\textproc{aggregate\_right}};
	\end{pgfonlayer}
\end{tikzpicture}
\end{adjustbox}

\caption{Visualization of an exemplary tree traversal as performed by \cref{alg:aggregate_once} to compute $\lift{\f}{\mathcal{M}}\bigl(\range{2}{13}{\V(t)}\bigr)$. Notice that $v_7$ need not be visited, as its contribution to the accumulated value is already part of $\liftlabel{\f}{\mathcal{M}}(v_8)$. Notice further that the traversal visits $v_{14}$ but ignores it, as $14$ lies outside the range.}

\label{fig:aggregate_once}
\end{figure*}

Overall, \cref{alg:aggregate_once} performs searches for two items in a search tree, along with some constant-time computations in each search step. If $t$ is balanced, the time complexity is thus in $\complexity{\log(\abs{\V(t)})}$. As the algorithm requires no dynamic memory allocation and is not recursive, its space complexity is in $\complexity{1}$.

Because associativity guarantees equal results regardless of the precise shape of the tree, implementations need not restrict themselves to binary trees. The algorithm can be extended to B-trees~\cite{bayer2002organization}, for example.

\subsection{Monoidal Fingerprints}

Now that we have characterized a general family of functions that admit efficient computation on ranges, we can turn back to the range-based set reconciliation approach. \Cref{protocol:rbsr} works by recursively testing fingerprints for equality. For our purposes,
we can define a fingerprint or hash function as follows:

\begin{definition}
A \defined{hash function} is a function $\fun{\h}{U}{D}$ with a finite codomain, such that, for randomly chosen $u \in U$ and $d \in D$, the probability that $\h(u) = d$ is roughly\footnote{To keep our focus on set reconciliation rather than being sidetracked by cryptography, we will for the most part keep arguments about probabilities qualitative rather than quantitative.} $\frac{1}{\abs{D}}$. $\h(u)$ is called the \defined{hash of $u$}, \defined{fingerprint of $u$} or \defined{digest of $u$}.
\end{definition}

To efficiently compute fingerprints for arbitrary ranges, we use \somewhatmorphisms{} $\lift{\f}{\mathcal{M}}$ that serve as hash functions from $\powerset{U}$. As $\lift{\f}{\mathcal{M}}(\set{u})$ is equal to $\f(u)$, $\f$ must itself already be a hash function. Typical hash functions map values to bit strings of a certain length, i.e., the codomain is $\set{0, 1}^k$ for some $k \in \N$. We will thus consider monoids whose elements can be represented by such bit strings.

A natural choice of the monoid universe is then $\range{0}{2^k}{\N}$, some simple monoidical operations on this universe include bitwise xor, addition modulo $2^k$, and multiplication modulo $2^k$. Of these three options, multiplication is the least suitable, because multiplying any number by $0$ yields $0$. Consequently, for every set containing an item $u$ with $\f(u) = 0$, the fingerprint of the set is $0$, which clearly violates the criterion that all possible values for fingerprints occur with equal probability.

The monoid operation preserves a good distribution of fingerprints if any given fingerprint can be obtained from any particular fingerprint by combining it with some third one, i.e., if, for every $x \in M$, $\lambda y.x \groupaddsym y$ is a bijection. Addition and xor satisfy this criterium, as does in fact every finite commutative group $\mathcal{G} = (G, \groupaddsym, -)$: for every $x, z \in M$ there exists $y \in M$ such that $x \groupaddsym y = z$, by choosing $y \defeq z \groupaddsym -x$, because then $x \groupaddsym y = x \groupaddsym z \groupaddsym (-x) = x \groupaddsym (-x) \groupaddsym z = z$. Hence, $\lambda y.x \groupaddsym y$ is surjective, and, because $G$ is finite, the function is also injective.

By using such a \somewhatmorphism{} $\lift{\f}{\mathcal{M}}$, we can efficiently implement range-based set reconciliation. A node stores its set in a monoid tree labeled by both $\liftlabel{\f}{\mathcal{M}}$ and $\liftlabel{\lambda x.1}{\mathcal{N}}$. On receiving a range fingerprint $\ifp{x}{y}{X_j}$, the node \peer{i} efficiently computes $\lift{\f}{\mathcal{M}}\bigl(\range{x}{y}{X_i}\bigr)$. If the fingerprints do not match, it computes $\lift{\lambda x.1}{\mathcal{N}}\bigl(\range{x}{y}{X_i}\bigr)$ to determine the number of items it has in the range, and uses this information for determining the sizes of the subranges to create. Finding the boundaries of those subranges amounts to looking up items by index in an order-statistic tree, and thus takes logarithmic time. All of these operations require only $\complexity{1}$ space.

Overall, the computations for processing a single range fingerprint for a local set of size $n_i$ thus take $\complexity{\log(n_i)}$ time. As a single message can contain $\complexity{n_{\triangle}}$ many range fingerprints, where $n_{\triangle}$ is the size of the symmetric difference of the sets to reconcile, the overall time complexity per communication round is in $\complexity{n_{\triangle} \cdot \log(n_i)}$.

\subsection{Ascending Intervals}

When computing fingerprints for several ranges, we can reduce the overall time complexity if the ranges are sorted by their lower boundaries in ascending order. We can accumulate labels while traversing from the lower boundary of each range to its upper boundary; then we traverse to the lower boundary of the next range, ready to process it.

The maximum distance between two vertices in a balanced tree on $n$ vertices is in $\complexity{\log(n)}$. Processing any individual range this way thus requires $\complexity{\log(n)}$ time, just like our previous approach. Notice however that, when traversing the tree for successive sorted ranges, every edge is traversed at most twice. The time complexity for traversing multiple ranges in sequence is thus at most in $\complexity{n}$. Using this approach, we can hence bound the time complexity for a single communication round by $\complexity{\min(n_i, n_{\triangle} \cdot \log(n_i))}$. This can result in a logarithmic speed-up compared to prior discussion of range-based set reconciliation (\cite{chen1999prototype}\cite{shang2017survey}).

\textproc{aggregate\_until} (\cref{alg:aggregate_ascending}) implements this traversal as a procedure that takes the boundaries of a single range and the vertex that stores the lower boundary as arguments, and returns both the aggregated monoidal value of the range, and the vertex that stores the least value that is greater than the upper boundary. This vertex can be used as the starting point for the next invocation of the procedure to find the lower boundary of the next range. If no such vertex exists, the procedure returns $\nil$ in its place, and the aggregated value for all following ranges is known to be $\neutraladd$.

Observe that the path from some lower boundary $x$ to some upper boundary $y$ consists of some (possibly zero) steps from $x$ toward the root, and then some (possibly zero) steps toward $y$. In order to compute this path in constant space and time per step, we augment the data structure by adding to each vertex $v$ a reference $\tp{v}$ to its parent ($nil$ for the root), and the largest value stored in its subtree $\tm{v}$. The traversal begins by following parent references until reaching the root of a subtree $t$ that contains a value greater than or equal to $y$ (\textproc{aggregate\_up}), which we can efficiently detect by comparing $\tm{t}$ against $y$. The successive downward traversal (\textproc{aggregate\_down}) for finding the least value above the range terminates upon reaching a vertex $t$ with $\tv{t} \succeq y$ whose left subtree is fully contained within the range, i.e., with $\tm{\tl{t}} \prec y$.

\begin{algorithm}
\caption{}\label{alg:aggregate_ascending}
\begin{algorithmic}[1]
\Require $x \prec y$
\Procedure{aggregate\_until}{$t, x, y$}
	\State $(acc, t) \gets \Call{aggregate\_up}{t, x, y}$
	\If{$t = \nil \lor \tv{t} \succeq y$}
		\State \return $(acc, t)$
	\Else
		\State \return $\Call{aggregate\_down}{\tr{t}, y, acc \groupaddsym \f(\tv{t})}$
	\EndIf
\EndProcedure

\Procedure{aggregate\_up}{$t, x, y$}
	\State $acc \gets \neutraladd$
	\While{$\tm{t} \prec y$}
		\If{$\tv{t} \succeq x$}
			\State $acc \gets acc \groupaddsym \f(\tv{t}) \groupaddsym \liftlabel{\f}{\mathcal{M}}(\tr{t})$
		\EndIf
		\If{$\tp{t} = \nil$}
			\State \return $(acc, \nil)$
		\Else
			\State $t \gets \tp{t}$
		\EndIf
	\EndWhile
	\State \return $(acc, t)$
\EndProcedure

\Procedure{aggregate\_down}{$t, y, acc$}
	\While{$t \neq \nil$}
		\If{$\tv{t} \prec y$}
			\State $acc \gets acc \groupaddsym \liftlabel{\f}{\mathcal{M}}(\tl{t}) \groupaddsym \f(\tv{t})$
			\State $t \gets \tr{t}$
		\ElsIf{$\tl{t} = \nil \lor \tm{\tl{t}} \prec y$}
			\State \return $(acc \groupaddsym \liftlabel{\f}{\mathcal{M}}(\tl{t}), t)$
		\Else
			\State $t \gets \tl{t}$
		\EndIf
	\EndWhile
	\State \return $(acc, \nil)$
\EndProcedure
\end{algorithmic}
\end{algorithm}

% algebra TODO
% Arbitrary predicates, queries, higher dimensional ranges TODO

\section{Adversarial Environments}\label{secure}

\Cref{protocol:rbsr} uses fingerprints of sets for probabilistic equality checking: we assume sets with equal fingerprints to be equal. Synchronization can thus become faulty if it involves unequal sets with equal fingerprints. If the universe of possible fingerprints is chosen large enough, and the distribution of fingerprints of randomly chosen sets is randomly distributed within that universe, the probability for this to occur becomes negligible.

Random distribution of input sets is however a very strong assumption. In this section, we examine how to mitigate a malicious adversary that can influence the sets to be fingerprinted, with the goal of causing fingerprint collisions and consequently triggering faulty behavior of the system.

\subsection{Impact of Hash Collisions}

We can generally distinguish between malicious actors in two different positions: those who can actively impact the contents of the data structure to be synchronized, and those who passively relay updates and need to search for a collision within the available data. As a set of size $n$ has $2^n$ subsets, if fingerprints are bit strings of length $k$, then by the pigeonhole principle a fingerprint collision can be found within any set of size at least $k + 1$.

An attack against the fingerprinting scheme by an active adversary can involve computing many fingerprints and adding the required items to the set once a collision has been found. Such an attack is not usable by the passive adversary. We will primarily focus on discussing active adversaries, as they are strictly more powerful than passive ones. Yet it should be kept in mind that passive adversaries can be more common in certain settings, particularly in peer-to-peer systems: if a node is interested in synchronizing a data structure, it probably trusts the source of the data, otherwise it would have little reason for expending resources on synchronization. The data may however be synchronized not with the original source but with completely untrusted nodes.

Fingerprint collisions result in parts of the data structure not being synchronized, so information is being withheld from one or both of the synchronizing nodes. When a malicious node synchronizes with an honest one, the malicious node can withhold arbitrary information by simply pretending not to have certain data, which does not require finding collisions at all.

So the cases in which a malicious node can do actual damage by finding a collision are those where it supplies data to two honest nodes such that these two nodes perform faulty synchronization amongst each other. Specifically: let $\mathcal{M}$ be a malicious node, $\mathcal{A}$ and $\mathcal{B}$ be honest nodes, then a successful attack consists of $\mathcal{M}$ crafting sets $X_A, X_B$ and sending these to $\mathcal{A}$ and $\mathcal{B}$ respectively, so that when $\mathcal{A}$ and $\mathcal{B}$ then run the synchronization protocol, they end up with distinct sets. A passive adversary does not craft $X_A, X_B$ but must find them as subsets of some set $X$ supplied by an honest node.

There are some qualitative arguments that even if an adversary finds a fingerprint collision, the impact is rather low. Let $S_A \subseteq X_A$ and $S_B \subseteq X_B$ be nonequal sets with the same fingerprint. To have any impact on the correctness of a particular protocol run, their two fingerprints need to actually be compared during that run. For that to happen, there have to be $x, y \in U$ such that $S_A = \range{x}{y}{X_A}$ and $S_B = \range{x}{y}{X_B}$.

If the adversary has found such sets, this provides still no guarantee that the range $\range{x}{y}{X_i}$ is being compared during the synchronization session of $\mathcal{A}$ and $\mathcal{B}$. In particular, there is no need for $\mathcal{A}$ and $\mathcal{B}$ to choose the range boundaries that occur in a protocol run deterministically. They can, for example, split ranges into equally-sized subranges first, but then randomly shift the range boundaries by a small number of items. This preserves a logarithmic number of communication rounds in the worst case, while also making it impossible for an attacker to make sure that a given hash collision will impact a given synchronization session. In order to minimize the probability that a particulr collision affects a given protocol run, the $b$ subrange boundaries can be chosen fully at random. The expected number of communication rounds is in $\complexity{\log(n)}$ with high probability, as it corresponds to the height of a randomly chosen $b$-complete tree, which can be expected to be within a constant factor of $\log_b(n)$~\cite{devroye1990height}.

Another factor mitigating the impact of an adversary finding fingerprint collisions is the communication with other, non-colluding nodes. A fourth party could send some $u \in U, x \preceq u \prec y$ to $\mathcal{A}$ or $\mathcal{B}$ before $\mathcal{A}$ and $\mathcal{B}$ synchronize, disrupting the collision.

Finally, in systems where nodes repeatedly synchronize with different other nodes, a single fingerprint collision in a single synchronization session would merely delay propagation of information rather than stop it completely. Peer-to-peer systems communicating on a random overlay network in particular fall into this category. A malicious actor with enough control over the communication of other nodes to guarantee a tangible benefit from fingerprint collisions can likely disrupt operation of the network more effectively by exercising that control than by sabotaging synchronization.

All of these arguments are however purely qualitative and should as such be taken into account with caution, they are not a substitute for quantitative cryptographic analysis. A strong attacker might be able to find many pairs of sets of colliding fingerprints, or many sets that all share the same fingerprint, and none of the above arguments consider these cases.

\subsection{Cryptographically Secure Fingerprints}

For stronger guarantees, we thus look at cryptographically secure fingerprint functions that make it computationally infeasible for an adversary to find inputs that lead to faulty synchronization.

A typical definition of cryptographically secure hash functions is the following~\cite{menezes2018handbook}:

\begin{definition}
A \defined{secure hash function} is a hash function $\fun{\h}{U}{D}$ that satisfies three additional properties:

\begin{description}
  \item[pre-image resistance:] Given $d \in D$, it is computationally infeasible to find a $u \in U$ such that $\h(u) = d$.
  \item[second pre-image resistance:] Given $u \in U$, it is computationally infeasible to find a $u' \in U, u' \neq u$ such that $\h(u) = \h(u')$.
  \item[collision resistance:] It is computationally infeasible to find $u, v \in U, u ~= v$ such that $\h(u) = \h(v)$.
\end{description}
\end{definition}

What do secure fingerprints for our sets look like? Since $\lift{\f}{\mathcal{M}}(\set{u}) = \f(u)$, $\f$ must necessarily be a secure hash function if $\lift{\f}{\mathcal{M}}$ is to be one. This alone is unfortunately not sufficient, as demonstrated by xor as the monoid function:~\cite{bellare1997new} shows how to reduce the problem of finding a collision to that of solving a system of linear equations over the finite field on two elements (in which xor is the additive operation), which can be done in cubic time.

We now present further monoids that have been studied in the construction of secure hash functions. The seminal~\cite{bellare1997new} considers secure hash functions for strings such that the hash of the concatenation of two strings can be efficiently computed from the hashes of the two strings. By considering sets as strings of their items in ascending order, the functions studied in~\cite{bellare1997new} can also be applied to our sets. After demonstrating the unsuitability of xor, the authors consider addition modulo the total number of possible hashes, and the multiplicative group $\Z_n^\ast$, the group yielded by multiplication modulo $n$ on the set $\set{x \in \range{0}{n}{\N} \mid \text{$x$ is coprime to $n$}}$.

They unify parts of their discussion by relating the hardness of finding collisions to solving the balance problem: in a commutative group $(G, \groupaddsym, \neutraladd)$, given a set of group elements $S = \set{s_1, s_2, \ldots, s_n}$, find disjoint, nonempty subsets $S_0 = \set{s_{0, 0}, s_{0, 1}, \ldots, s_{0, k}} \subseteq S, S_1 = \set{s_{1, 0}, s_{1, 1}, \ldots, s_{1, l}} \subseteq S$ such that $s_{0, 0} \groupaddsym s_{0, 1}  \groupaddsym \ldots  \groupaddsym s_{0, k} = s_{1, 0}  \groupaddsym s_{1, 1}  \groupaddsym \ldots  \groupaddsym s_{1, l}$. They then reduce the hardness of the balance problem to other problems.

For addition, the balance problem is as hard as subset sum, which was at the time of publication conjectured to be sufficiently hard. Wagner showed however in~\cite{wagner2002generalized} how to solve the balance problem in subexponential time for addition.~ \cite{mihajloska2015reviving} suggests addition for combining SHA-3~\cite{dworkin2015sha} digests, and proposes using fingerprints of length between $2688$ and $4160$ or $6528$ to $16512$ bits to achieve security levels of $128$ or $256$ bit respectively against Wagner's attack.~\cite{lyubashevsky2005parity} gives an improvement over Wagner's attack finding collisions in $\complexity{2^{n^\epsilon}}$ for arbitrary $\epsilon < 1$, further weakening addition as a choice of monoid operation.

For multiplication, the balance problem is as hard as the discrete logarithm problem in the group. This is a more ``traditional'' hardness assumption than subset sum; there are groups for which no efficient algorithm is known. The main drawback is that multiplication is less efficient to compute than addition.~\cite{stanton2010fastad} includes a comparison between the performance of addition and multiplication for incremental hashing; the additive hash outperforms the multiplicative one by two orders of magnitude, even though the additive hashes use longer digests to account for Wagner's attack.

When fingerprints are frequently sent over the network, longer computation times might be preferable over longer hashes. Fingerprints based on multiplication nevertheless need larger digests than traditional, non-incremental hash functions;~\cite{maitin2017elliptic} suggests fingerprints of $3200$ bit to achieve $128$ bit security.

\cite{bellare1997new} also proposes a fourth monoid based on lattices. \cite{lewi2019securing} give a specific instantiation providing 200 bits of security with fingerprints of size $16 \cdot 1024 = 16384$ bit.

\cite{clarke2003incremental} identifies the hash functions of~\cite{bellare1997new} to be \defined{multiset homomorphic}:

\begin{definition}[Monoid Homomorphism]
Let $\mathcal{U}_0 \defeq (U_0, \groupaddsym_0, \neutraladd_0)$ and $\mathcal{U}_1 \defeq (U_1, \groupaddsym_1, \neutraladd_1)$ be monoids, and let $\fun{\f}{U_0}{U_1}$.

We call $\f$ a \defined{monoid homomorphism from $\mathcal{U}_0$ to $\mathcal{U}_1$} if for all $x, y \in U_0$ we have $\f(x \groupaddsym_0 y) = \f(x) \groupaddsym_1 \f(y)$.
\end{definition}

\begin{definition}[Multiset Homomorphic Hash Function]
Let $\mathcal{S} \defeq (\N^U, \cup, \emptyset)$ be the monoid of multisets over the universe $U$ under union, $\mathcal{M} \defeq (M, \groupaddsym, \neutraladd)$ a monoid, and $\fun{\f}{\N^U}{M}$.

We call $\f$ a \defined{multiset homomorphic hash function} if $\f$ is a hash function and a monoid homomorphism from $\mathcal{S}$ to $\mathcal{M}$.
\end{definition}

Being a monoid homomorphism from $\mathcal{S}$ to $\mathcal{M}$ is a strictly stronger criterium than being a \somewhatmorphism{}. Hence, every multiset homomorphic hash function is suitable for our purposes.

Beyond the multiset homomorphic hash functions introduced in~\cite{bellare1997new}, \cite{cathalo2009comparing} provides a construction based on RSA, and~\cite{maitin2017elliptic} provides an efficient construction based on elliptic curves.

Because multiset union is commutative, so is necessarily any multiset homomorphic hash function. We do not require commutativity for our fingerprints however. A typical associative but not commutative operation is matrix multiplication. Study of a family of hash functions based on multiplication of invertible matrices was initiated in \cite{zemor1991hash}. The security of these hash functions is related to solving hard graph problems on the Cayley graph of the matrix multiplication group. \cite{petit2011rubik} gives an overview about the general principles and the security aspects of Cayley hash functions.

While~\cite{tillich1994hashing}, an improvement over the originally proposed scheme, has been successfully attacked in~\cite{grassl2011cryptanalysis} and~\cite{petit2010preimages}, no attacks are known for several modifications such as~\cite{petit2009graph}\cite{bromberg2017navigating}\cite{sosnovski2016cayley}; and~\cite{mullan2016text} shows random self-reducibility for Cayley hash functions.

Aside from Cayley hashes, we are not aware of any non-commutative monoids used for hashing. Notice that hash functions based on non-commutative groups (such as Cayley hashes) still have more structure than we need, as we don't require existence of inverse elements. Suitable hash functions can thus be located in a more general design space than studied in any literature we know of.

Regardless of the choice of monoid, the reconciliation protocol can exchange hashes of fingerprints rather than exchanging the fingerprints directly. This allows us to use large fingerprints to achieve security (e.g., using bitwise additions on long bitstrings as the monoid), while still transmitting only a small number of bits over the wire. The large fingerprints do however increase the space consumption of the monoid tree. Whether a computationally expensive monoid operation on small bitstrings outperforms a cheaper operation on larger bitstrings is hence not obvious and requires benchmarking to make an informed choice.

\section{Pseudorandom Trees}
\label{randomization}

Monoidal fingerprints allow efficient deterministic computation of fingerprints, but at the price of having to rely on some non-standard cryptographic primitives. A more established cryptographic construction is that of Merkle trees~\cite{merkle1989certified} and related data structures, all based on hashing the concatenation of child hashes. There is a significant body of work in the context of authenticated data structures based on these constructions, from rather simple balanced binary trees~\cite{naor2000certificate} to arbitrary acyclic directed graphs~\cite{martel2004general}.

The Merkle construction is however not associative when using a regular secure hash function $\h$, as $\h(\h(\h(a) \concatenate \h(b)) \concatenate \h(c)) = \h(\h(a) \concatenate \h(\h(b) \concatenate \h(c)))$ would constitute a violation of collision resistance. Different tree representations of the same set thus lead to different fingerprints. So in order to use this construction, we must ensure a unique search tree representation for every set.

\cite{uniquerepresentation} shows that deterministic search trees with unique shapes require $\complexity{\sqrt{n}}$ time for insertions and deletions in the worst case, making them unsuitable for our use case. A natural extension is to look at randomized data structures that define a unique representation which allows modification and search in $\complexity{\log(n)}$ with high probability.

Such data structures have been thoroughly researched in the field of \defined{history-independent data structures} --- intuitively speaking, data structures whose bitlevel representations do not leak information about their construction sequence. This requirement has been shown in~\cite{hartline2005characterizing} to be equivalent to having a unique representation. For more background, we refer to the excellent introduction of~\cite{bender2016anti}.

In the following, we give a construction based on treaps, and define Merkle-like fingerprinting schemes for treaps, such that the fingerprints of arbitrary ranges can be computed in logarithmic time with high probability. As all nodes need to arrive at the same representation, no true randomness can be involved. Instead, all random decisions are based on pseudorandom bits derived from the data itself.

This scheme allows for cryptographically secure fingerprints whose collision resistance does not depend on uncommon cryptographic building blocks. The price to pay is that computing fingerprints can take linear time in the worst case. Note however that the communication complexity of any synchronization protocol using these fingerprints is completely unaffected by the randomized computation time, as nodes can still partition ranges into subranges of equal size. In this regard, our approach is more robust than that of~\cite{auvolat2019merkle}, in which the number of communication rounds depends on the height of a pseudorandom tree.

Unfortunately, adversaries can efficiently create data sets for which the randomized solutions degrade to linear performance, thus enabling denial-of-service attacks as the cost of computing the fingerprints during a protocol run can be made to dominate the communication cost. Passive adversaries however cannot effectively attack these fingerprints: even though they can technically only forward those parts of the data structure that have degraded performance relative to the number of items in those parts, they would still increase the absolute resource usage of their victims if they simply forwarded all data.

Proponents of randomized data structures also claim that they are significantly easier to implement than balanced search trees (see~\cite{seidel1996randomized} or~\cite{pugh1990skip}, for example), making it worthwhile to consider them even in trusted environments, in which their expected logarithmic update complexities suffice.

\subsection{Pseudorandom Treaps}

A \defined{treap}~\cite{seidel1996randomized} is a search tree in which every vertex has an associated \defined{priority} from a totally ordered set, and in which the priorities of the vertices form a heap:

\begin{definition}
Let $U$ be a set, $\prec$ a linear order on $U$, $P$ a set, $\leq$ a linear order on $P$, $\fun{\priority}{U}{P}$, $V \subseteq U$, and $T$ a binary search tree on $V$.

Then we call $T$ a treap if for all $v \in V$ with parent $p$ it holds that $\priority(v) \leq \priority(p)$.
\end{definition}

From the properties of treaps shown in~\cite{seidel1996randomized} we will rely on the following:

\begin{itemize}
  \item If the priorities of all vertices are pairwise unequal, then there is exactly one treap on the vertex set, which is equal to the search tree obtained by successively inserting items in decreasing priority without any balancing.
  \item If the priorities are distributed uniformly at random, the height of the treap is expected to be logarithmic in the number of vertices.
  \item Inserting or deleting an item while maintaining the treap properties can be done in time proportional to the height of the treap.
\end{itemize}

In the following, we will fix some cryptographically secure hash function $\fun{\p}{U}{\set{0, 1}^k}$, and define $\priority(u) \defeq \p(u)$, using the numeric order on $\set{0, 1}^k$ for comparing priorities. Since $\p$ is collision resistant, we can assume the resulting treaps to be unique, and since the output of a secure hash function is indistinguishable from a random mapping, we can assume the resulting treaps to have expected logarithmic height.

Treaps store items in every vertex, whereas Merkle trees only store items in their leaves. \cite{buldas2002eliminating} proposes the following natural generalization and proves that the labels are collision free if the underlying hash function $\h$ is collision free:

\begin{align*}
\tlabel{\h}(t) \defeq&\\
\begin{cases}
\h(\tv{t}), &  \text{$\tl{t} = \nil$ and $\tr{t} = \nil$}\\
\tlabel{\h}(\tlabel{\h}(\tl{t} \concatenate \h(\tv{t}) \concatenate \tlabel{\h}(\tr{t}))) & \text{$\tl{t} \neq \nil$ and $\tr{t} \neq \nil$}\\
\tlabel{\h}(\tlabel{\h}(\tl{t}) \concatenate \tlabel{\h}(\tv{t})) & \text{$\tl{t} \neq \nil$ and $\tr{t} = \nil$}\\
\tlabel{\h}(\tlabel{\h}(v) \concatenate \tlabel{\h}(c_{>})) & \text{$\tl{t} = \nil$ and $\tr{t} \neq \nil$}\\
\end{cases}
\end{align*}

%   \[
%    \tlabel{\h}(t) \defeq \begin{cases}
% \h(\tv{t}), &  \text{$\tl{t} = \nil$ and $\tr{t} = \nil$}\\
% \tlabel{\h}(\tlabel{\h}(\tl{t} \concatenate \h(\tv{t}) \concatenate \tlabel{\h}(\tr{t}))) & \text{$\tl{t} \neq \nil$ and $\tr{t} \neq \nil$}\\
% \tlabel{\h}(\tlabel{\h}(\tl{t}) \concatenate \tlabel{\h}(\tv{t})) & \text{$\tl{t} \neq \nil$ and $\tr{t} = \nil$}\\
% \tlabel{\h}(\tlabel{\h}(v) \concatenate \tlabel{\h}(c_{>})) & \text{$\tl{t} = \nil$ and $\tr{t} \neq \nil$}\\

% \end{cases}
%   \]

We will call a treap labeled according to this scheme a \defined{Merkle treap}. Since Merkle treaps are unique, we can define the fingerprint of some set as the root label of the Merkle treap on that set if it is nonempty, or the hash of the empty set otherwise.

\subsection{Subsets}

Now that we have defined the treap-based fingerprint of a set, we need a way of efficiently computing the fingerprint of any range $\range{x}{y}{S}$ given the Merkle treap of the set $S$. We can do so by doing the same tree traversal as in \cref{alg:aggregate_once}, but fully ignoring any items outside the range; this yields \cref{alg:merkle}.

Notice that this algorithm is not specific to treaps, it works for any search tree that is labeled via $\tlabel{\h}$. For simplicity of presentation, we give a recursive formulation. An iterative version that can operate in constant space needs to first do the same traversal down the tree as the recursive version (but without performing any computations), before then following parent pointers back up the tree and computing the hashes along the way.

\begin{algorithm}
	\caption{Computing $\tlabel{\h}\bigl(\range{x}{y}{\V(t)}\bigr)$.}\label{alg:merkle}
	\begin{algorithmic}[1]
	\Require $x \preceq y, t \neq \nil$
	\Procedure{aggregate\_range}{$t, x, y$}
		\If{$x = y$}
			\State \return $\tv{t}$
		\EndIf
		\State $init \gets \Call{find\_initial}{t, x, y}$
		\If{$init = \nil$}
			\State \return $\h("")$
		\Else
			\State $h_{left} = \Call{aggregate\_left}{\tl{init}, x}$
			\State $h_{right} = \Call{aggregate\_right}{\tr{init}, y}$
			\State \return $\Call{combine}{h_{left}, \h(\tv{init}), h_{right}}$
		\EndIf
	\EndProcedure

	\Comment{Helper function for handling $\nil$ subtrees}
	\Procedure{combine}{$h_0, h_1, h_2$}
		\If{$h_0 = \h("") \land h_2 = \h("")$}
			\State \return $h_1$
		\ElsIf{$h_0 \neq \h("") \land h_2 = \h("")$}
			\State \return $\h(h_0 \cdot h_1)$
		\ElsIf{$h_0 = \h("") \land h_2 \neq \h("")$}
			\State \return $\h(h_1 \cdot h_2)$
		\ElsIf{$h_0 \neq \h("") \land h_2 \neq \h("")$}
			\State \return $\h(h_0 \cdot h_1 \cdot h_2)$
		\EndIf
	\EndProcedure
	
	\Procedure{aggregate\_left}{$t, x$}
		\If{$t = \nil$}
			\State \return $\h("")$
		\ElsIf{$\tv{t} \prec x$}
			\State \return $\Call{aggregate\_left}{\tr{t}, x}$
		\ElsIf{$\tv{t} = x$}
			\State \return $\Call{combine}{\h(""), \h(\tv{t}), \tlabel{\h}(\tr{t})}$
		\Else
			\State $h_{left} = \Call{aggregate\_left}{\tl{t}, x}$
			\State \return $\Call{combine}{h_{left}, \h(\tv{t}), \tlabel{\h}(\tr{t})}$
		\EndIf
	\EndProcedure

	\Procedure{aggregate\_right}{$t, y$}
		\If{$t = \nil$}
			\State \return $\h("")$
		\ElsIf{$\tv{t} \prec y$}
			\State $h_{right} = \Call{aggregate\_right}{\tr{t}, y}$
			\State \return $\Call{combine}{\tlabel{\h}(\tl{t}), \h(\tv{t}), h_{right}}$
		\ElsIf{$\tv{t} = x$}
			\State \return $\Call{combine}{\tlabel{\h}(\tl{t}), \h(\tv{t}), \h("")}$
		\Else
			\State \return $\Call{aggregate\_right}{\tl{t}, y}$
		\EndIf
	\EndProcedure
	\end{algorithmic}
\end{algorithm}

One can further modify \cref{alg:merkle} to efficiently traverse successive ascending ranges analogously to how \cref{alg:aggregate_ascending} arises from \cref{alg:aggregate_once}. The complexity bounds of (iterative versions of) these algorithms are the same as those for \cref{alg:aggregate_once} and \cref{alg:aggregate_ascending} respectively.

Just as with balanced search trees, a \textit{binary} treap might be the most natural formulation, but not the most efficient one on real hardware. \cite{golovin2009b} describes B-treaps, which are treaps of higher degree. Generalizing the Merkle labeling to k-ary trees is straightforward, compare~\cite{auvolat2019merkle}. If such an optimized realization is chosen, the need to map between the actual data layout and the shape of the treap which defines the fingerprints complicates the implementation, in particular compared to monoidal fingerprints which naturally abstract over tree implementation details.

Beyond treaps there are other randomized search trees. \cite{pugh1989incremental} suggests hash tries for fingerprinting sets, and our mechanism for computing fingerprints of ranges is compatible with their approach. The reason we opted for treaps is that the pseudorandom selection of the tree shape is completely decoupled from the ordering over the items, whereas in a hash trie the ordering follows from the tree shape (or vice versa, depending on the viewpoint). \cite{auvolat2019merkle} also gives a hash trie-like construction.

There exists also some work on deterministic unique tree representations that stay efficient if the number of items is either a particularly small or a particularly large fraction of the size of the universe, see~\cite{sundar1994unique} for both a specific approach and more related work. These approaches could also be adapted for our purposes, but since they pose restrictions on the number of items that can be stored, they are less flexible than a treap-based solution.

\subsection{Adversarial Treap Construction}

Pseudorandom schemes can rely on well-known, conventional, non-associative hash functions. While Merkle-treaps are collision-resistant, they open up another angle of attack: a malicious party can construct unbalanced treaps to make the cost of computing fingerprints super-logarithmic in the number of items, leading to a denial-of-service attack as the cost of computing the fingerprints during a protocol around dominates the communication cost.

There is a fairly straightforward way of creating treaps on $n$ vertices of height $n$ in expected $\complexity{n^2}$ time and $\complexity{1}$ space, making treaps unsuitable for an adversarial environment. A sequence $(u_0, u_1, \ldots, u_{n - 1})$ of items such that the treap on these items has height $n$ can be computed by successively choosing arbitrary but always increasing with respect to $\preceq$ items $p_j$ and letting $u_i \defeq p_j$ if $\floor{i \cdot \frac{2^k}{n}} \leq \priority(p_j) < \floor{(i + 1) \cdot \frac{2^k}{n}}$ (recall that $k$ is the number of bits of a hash digest). The probability for the priority of an arbitrary item to fall within the desired range is $\frac{1}{n}$, so finding one takes expected $\complexity{n}$ time, finding a sequence of $n$ of these accordingly takes expected $\complexity{n^2}$ time. Similar techniques can be used to create degenerate hash tries as well.

\section{Conclusion}\label{conclusion}

We consider range-based set reconciliation to be an important alternative to the constant-communication-round alternatives that make up the bulk of research on set reconciliation. Despite its comparative simplicity, its profile of complexity guarantees is not strictly worse (nor better) than that of any other approach. While the logarithmic number of communication rounds is a significant drawback, no other approach achieves computational complexity proportional to only the size of the symmetric difference. The option of allowing one endpoint to perform reconciliation with a constant amount of working memory is unique as well. These two factors can allow reconciliation in resource-constrained settings in which no other approach can function. Furthermore, the option of using a more sophisticated reconciliation procedure once ranges have become small enough can make it worth a consideration in almost any reconciliation scenario.

\section{Acknowledgments}

%Albus Dumbledore (anonymized for peer review)
Jan Winkelmann
 contributed the idea of hashing monoid values before transmission in order to use large monoid universes for secure fingerprints without affecting the number of transmitted bits.

\bibliographystyle{alphaurl}
\bibliography{main}
\end{document}